\RequirePackage{lineno}
\documentclass[aps,prd,preprint,superscriptaddress]{revtex4}
\usepackage{graphicx}

\newcommand{\met}{$E_T\hspace{-1.1em}/$\hspace{0.7em}}

\begin{document}


\title{Two-particle momentum correlations in jets produced in $p\bar p$ collisions at $\sqrt{s}=1.96$ TeV}


\affiliation{}


\date{\today}

\begin{abstract}
We present the first measurement of two-particle momentum correlations in jets produced in $p\bar p$ collisions at $\sqrt{s}=1.96$ TeV. Results are obtained for charged particles within a restricted cone with an opening angle of 0.5 radians around the jet axis and for events with dijet masses between 66 and 563 GeV/c$^{2}$. A comparison of the experimental data to theoretical predictions obtained for partons within the framework of resummed perturbative QCD in the next-to-leading log approximation (NLLA) shows that the parton momentum correlations survive the hadronization stage of jet fragmentation, giving further support to the hypothesis of local parton-hadron duality. The extracted value of the NLLA parton shower cutoff scale $Q_\mathit{eff}$ set equal to $\Lambda_\mathit{QCD}$ is found to be  $(1.4^{+0.9}_{-0.7})\times 100$ MeV. 
\end{abstract}


\affiliation{Institute of Physics, Academia Sinica, Taipei, Taiwan 11529, Republic of China} 
\affiliation{Argonne National Laboratory, Argonne, Illinois 60439} 
\affiliation{Institut de Fisica d'Altes Energies, Universitat Autonoma de Barcelona, E-08193, Bellaterra (Barcelona), Spain} 
\affiliation{Baylor University, Waco, Texas  76798} 
\affiliation{Istituto Nazionale di Fisica Nucleare, University of Bologna, I-40127 Bologna, Italy} 
\affiliation{Brandeis University, Waltham, Massachusetts 02254} 
\affiliation{University of California, Davis, Davis, California  95616} 
\affiliation{University of California, Los Angeles, Los Angeles, California  90024} 
\affiliation{University of California, San Diego, La Jolla, California  92093} 
\affiliation{University of California, Santa Barbara, Santa Barbara, California 93106} 
\affiliation{Instituto de Fisica de Cantabria, CSIC-University of Cantabria, 39005 Santander, Spain} 
\affiliation{Carnegie Mellon University, Pittsburgh, PA  15213} 
\affiliation{Enrico Fermi Institute, University of Chicago, Chicago, Illinois 60637} 
\affiliation{Comenius University, 842 48 Bratislava, Slovakia; Institute of Experimental Physics, 040 01 Kosice, Slovakia} 
\affiliation{Joint Institute for Nuclear Research, RU-141980 Dubna, Russia} 
\affiliation{Duke University, Durham, North Carolina  27708} 
\affiliation{Fermi National Accelerator Laboratory, Batavia, Illinois 60510} 
\affiliation{University of Florida, Gainesville, Florida  32611} 
\affiliation{Laboratori Nazionali di Frascati, Istituto Nazionale di Fisica Nucleare, I-00044 Frascati, Italy} 
\affiliation{University of Geneva, CH-1211 Geneva 4, Switzerland} 
\affiliation{Glasgow University, Glasgow G12 8QQ, United Kingdom} 
\affiliation{Harvard University, Cambridge, Massachusetts 02138} 
\affiliation{Division of High Energy Physics, Department of Physics, University of Helsinki and Helsinki Institute of Physics, FIN-00014, Helsinki, Finland} 
\affiliation{University of Illinois, Urbana, Illinois 61801} 
\affiliation{The Johns Hopkins University, Baltimore, Maryland 21218} 
\affiliation{Institut f\"{u}r Experimentelle Kernphysik, Universit\"{a}t Karlsruhe, 76128 Karlsruhe, Germany} 
\affiliation{Center for High Energy Physics: Kyungpook National University, Daegu 702-701, Korea; Seoul National University, Seoul 151-742, Korea; Sungkyunkwan University, Suwon 440-746, Korea; Korea Institute of Science and Technology Information, Daejeon, 305-806, Korea; Chonnam National University, Gwangju, 500-757, Korea} 
\affiliation{Ernest Orlando Lawrence Berkeley National Laboratory, Berkeley, California 94720} 
\affiliation{University of Liverpool, Liverpool L69 7ZE, United Kingdom} 
\affiliation{University College London, London WC1E 6BT, United Kingdom} 
\affiliation{Centro de Investigaciones Energeticas Medioambientales y Tecnologicas, E-28040 Madrid, Spain} 
\affiliation{Massachusetts Institute of Technology, Cambridge, Massachusetts  02139} 
\affiliation{Institute of Particle Physics: McGill University, Montr\'{e}al, Canada H3A~2T8; and University of Toronto, Toronto, Canada M5S~1A7} 
\affiliation{University of Michigan, Ann Arbor, Michigan 48109} 
\affiliation{Michigan State University, East Lansing, Michigan  48824} 
\affiliation{University of New Mexico, Albuquerque, New Mexico 87131} 
\affiliation{Northwestern University, Evanston, Illinois  60208} 
\affiliation{The Ohio State University, Columbus, Ohio  43210} 
\affiliation{Okayama University, Okayama 700-8530, Japan} 
\affiliation{Osaka City University, Osaka 588, Japan} 
\affiliation{University of Oxford, Oxford OX1 3RH, United Kingdom} 
\affiliation{University of Padova, Istituto Nazionale di Fisica Nucleare, Sezione di Padova-Trento, I-35131 Padova, Italy} 
\affiliation{LPNHE, Universite Pierre et Marie Curie/IN2P3-CNRS, UMR7585, Paris, F-75252 France} 
\affiliation{University of Pennsylvania, Philadelphia, Pennsylvania 19104} 
\affiliation{Istituto Nazionale di Fisica Nucleare Pisa, Universities of Pisa, Siena and Scuola Normale Superiore, I-56127 Pisa, Italy} 
\affiliation{University of Pittsburgh, Pittsburgh, Pennsylvania 15260} 
\affiliation{Purdue University, West Lafayette, Indiana 47907} 
\affiliation{University of Rochester, Rochester, New York 14627} 
\affiliation{The Rockefeller University, New York, New York 10021} 
\affiliation{Istituto Nazionale di Fisica Nucleare, Sezione di Roma 1, University of Rome ``La Sapienza," I-00185 Roma, Italy} 
\affiliation{Rutgers University, Piscataway, New Jersey 08855} 
\affiliation{Texas A\&M University, College Station, Texas 77843} 
\affiliation{Istituto Nazionale di Fisica Nucleare, University of Trieste/\ Udine, Italy} 
\affiliation{University of Tsukuba, Tsukuba, Ibaraki 305, Japan} 
\affiliation{Tufts University, Medford, Massachusetts 02155} 
\affiliation{Waseda University, Tokyo 169, Japan} 
\affiliation{Wayne State University, Detroit, Michigan  48201} 
\affiliation{University of Wisconsin, Madison, Wisconsin 53706} 
\affiliation{Yale University, New Haven, Connecticut 06520} 
\author{T.~Aaltonen}
\affiliation{Division of High Energy Physics, Department of Physics, University of Helsinki and Helsinki Institute of Physics, FIN-00014, Helsinki, Finland}
\author{J.~Adelman}
\affiliation{Enrico Fermi Institute, University of Chicago, Chicago, Illinois 60637}
\author{T.~Akimoto}
\affiliation{University of Tsukuba, Tsukuba, Ibaraki 305, Japan}
\author{M.G.~Albrow}
\affiliation{Fermi National Accelerator Laboratory, Batavia, Illinois 60510}
\author{B.~\'{A}lvarez~Gonz\'{a}lez}
\affiliation{Instituto de Fisica de Cantabria, CSIC-University of Cantabria, 39005 Santander, Spain}
\author{S.~Amerio}
\affiliation{University of Padova, Istituto Nazionale di Fisica Nucleare, Sezione di Padova-Trento, I-35131 Padova, Italy}
\author{D.~Amidei}
\affiliation{University of Michigan, Ann Arbor, Michigan 48109}
\author{A.~Anastassov}
\affiliation{Rutgers University, Piscataway, New Jersey 08855}
\author{A.~Annovi}
\affiliation{Laboratori Nazionali di Frascati, Istituto Nazionale di Fisica Nucleare, I-00044 Frascati, Italy}
\author{J.~Antos}
\affiliation{Comenius University, 842 48 Bratislava, Slovakia; Institute of Experimental Physics, 040 01 Kosice, Slovakia}
% ...
\collaboration{CDF Collaboration\footnote{With visitors from $^a$University of Athens, 15784 Athens, Greece, 
$^b$Chinese Academy of Sciences, Beijing 100864, China, 
$^c$University of Bristol, Bristol BS8 1TL, United Kingdom, 
$^d$University Libre de Bruxelles, B-1050 Brussels, Belgium, 
$^e$University of California Irvine, Irvine, CA  92697, 
$^f$University of California Santa Cruz, Santa Cruz, CA  95064, 
$^g$Cornell University, Ithaca, NY  14853, 
$^h$University of Cyprus, Nicosia CY-1678, Cyprus, 
$^i$University College Dublin, Dublin 4, Ireland, 
$^j$University of Edinburgh, Edinburgh EH9 3JZ, United Kingdom, 
$^k$University of Heidelberg, D-69120 Heidelberg, Germany, 
$^l$Universidad Iberoamericana, Mexico D.F., Mexico, 
$^m$University of Manchester, Manchester M13 9PL, England, 
$^n$Nagasaki Institute of Applied Science, Nagasaki, Japan, 
$^o$University de Oviedo, E-33007 Oviedo, Spain, 
$^p$Queen Mary, University of London, London, E1 4NS, England, 
$^q$Texas Tech University, Lubbock, TX  79409, 
$^r$IFIC(CSIC-Universitat de Valencia), 46071 Valencia, Spain. 
}}
\noaffiliation

\maketitle


\section{Introduction}
 The goal of this analysis is to measure the two-particle momentum correlation in jets, study its dependence on jet energy, and compare the results to analytical predictions of the next-to-leading log approximation (NLLA) {\cite{NLLA}}.

	The evolution of jets is driven by the emission of gluons with very small transverse momenta with respect to the jet axis, i.e. less than $1$ GeV/c. The theoretical predictions, which are compared with the results of this measurement, are based on NLLA calculations supplemented with the hypothesis of local parton-hadron duality (LPHD) {\cite{lphd}}. NLLA provides an analytical description of parton shower formation, while LPHD states that the hadronization process takes place locally and, therefore, properties of partons and hadrons are closely related. Detailed studies of jet fragmentation allow one to better understand the relative roles of perturbative parton showering and non-perturbative hadronization in shaping the main jet characteristics. Past experimental studies of inclusive distributions of particles in jets in $e^{+}e^{-}$ {\cite{Early}} and $p\bar p$ {\cite{SafonovPRL,SafonovPRD}} collisions have shown good agreement with theoretical predictions, suggesting that the perturbative QCD (pQCD) stage must dominate jet formation, and the role of the non-perturbative stage is reduced to converting final partons into hadrons without significantly affecting their multiplicities and momenta. This paper addresses the question of whether more subtle effects, such as momentum correlation, also survive hadronization. The predictions for the parton momentum correlations at the level of NLLA precision were first obtained by C.P.~Fong and B.R.~Webber in {\cite{Webber}} and recently recalculated in the modified leading log approximation (MLLA) framework by R.~Perez-Ramos {\cite{Redamy}}. These pQCD-driven correlations extend over a large range of parton momenta differences and should not be confused with phenomenological Bose-Einstein correlations \cite{BEcorr} present only for parton momenta differences up to $200$ MeV.

Until now, the two-particle momentum correlations were studied only by the OPAL collaboration in $e^{+}e^{-}$ collisions at a center of mass energy of $\sim91$ GeV {\cite{OPALcorr}}. Charged particles in the full experimentally accessible solid angle were used in the OPAL analysis. This made it possible for OPAL to ignore some effects of jet reconstruction, but the correlations were measured over a larger cone radius than can be reliably treated theoretically. Strictly speaking, the theory describes parton shower development only within a small opening angle $\theta_{c}$ around the jet axis, so that $\tan\theta_{c} \sim\theta_{c}$. The OPAL measured distributions showed a pattern in qualitative agreement with theoretical predictions, but the values of the parton shower cutoff $Q_\mathit{eff}$ \cite{NLLA} extracted from the fit of the correlation distributions were inconsistent with the $Q_\mathit{eff}$ extracted from fits of the inclusive momentum distributions {\cite{OPALmom}}.

In this paper, we report a measurement of the two-particle momentum correlation for charged particles in events with dijet invariant masses in the range 66--563 GeV/c$^{2}$. Momentum correlation distributions are measured for charged particles in restricted cones with an opening angle of $\theta_{c}=0.5$ radians around the jet axis. Events were produced at the Tevatron collider in $p\bar p$ collisions at a center of mass energy of 1.96 TeV and were recorded by the CDF II detector. The total integrated luminosity is 385 pb$^{-1}$.

The data sample consists of events with an expected mixture of quark and gluon jets. In order to compare experimental results with theoretical predictions, the analysis is carried out in the center-of-mass system of the two jets. The data are divided into seven bins according to the value of dijet invariant mass, which allows us to explore the dependence of correlation parameters on energy. 

The data are fit to NLLA analytical functions and the value of the parton shower cutoff $Q_\mathit{eff}$ is extracted. The correlations observed in data are compared to Monte Carlo predictions by the {\sc pythia} tune A {\cite{Pythia,PythiaTuneA1}} and {\sc herwig}~6.5 {\cite{Herwig}} event generators.

\section{Theory}
	\subsection{Next-to-leading log approximation}
	NLLA allows a perturbative calculation of the parton shower by keeping all terms of order $\alpha_{s}^{n}\ln^{2n}(E_\mathit{jet})$ and $\alpha_{s}^{n}\ln^{2n-1}(E_\mathit{jet})$ at all orders $n$ of perturbation theory. In these terms $\alpha_{s}$ is the strong coupling constant and $E_\mathit{jet}$ is the jet energy. Most of the particles in jets have $k_T < 1$ GeV/c \cite{SafonovPRD}, where $k_T$ is the transverse momentum with respect to the jet axis. Therefore, in order to successfully describe jet fragmentation, a theoretical model must be able to handle parton emissions at such low transverse momenta scales.

	In NLLA the requirement that parton $k_{T}>Q_\mathit{cutoff}$, for a sufficiently high cutoff scale $Q_\mathit{cutoff}$ (typically a few GeV/c), ensures that only partons in the perturbative region are included in the calculation. After the resummation the value of the parameter $Q_\mathit{cutoff}$ can be lowered to the value of $\Lambda_{QCD}$. Lowering the parameter $Q_\mathit{cutoff}$ is equivalent to including softer partons in the description of the model. Setting $Q_\mathit{cutoff}$ to its lowest value, $\Lambda_{QCD}$, maximizes the range of applicability of the model. The phenomenological scale replacing the two initial parameters $Q_\mathit{cutoff}$ and $\Lambda_{QCD}$ is conventionally called $Q_\mathit{eff}$. In theoretical calculations final expressions for the observables of interest are usually functions of $\tau=\ln(Q/Q_\mathit{eff})$, where $Q=E_\mathit{jet}\theta_{c}$ is the so-called jet hardness. These final expressions can be expanded in powers of $\tau$. The Fong-Webber approach {\cite{Webber}} keeps only terms that are fully controlled within the precision of the calculation, i.e. neglects all terms of order $\alpha_{s}^{n}\ln^{2n-2}(E_\mathit{jet})$ and higher. The Perez-Ramos approach {\cite{Redamy}} partially includes higher-order terms.   

	\begin{figure}
	\includegraphics[width=3.2in]
	{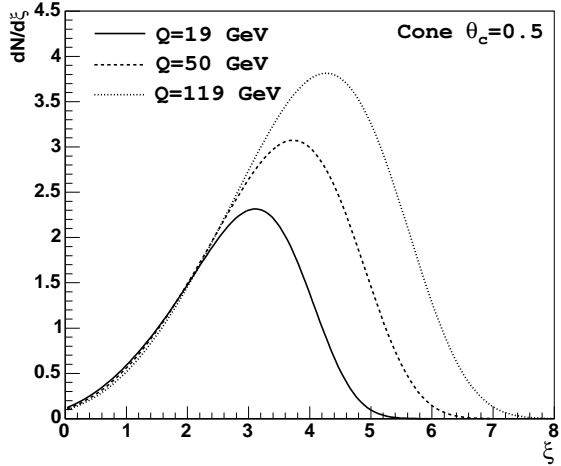}
	\caption{NLLA inclusive parton momentum distributions for $Q=E_\mathit{jet}\theta_{c}=19$, $50$, and $120$ GeV and $Q_\mathit{eff}=230$ MeV calculated according to {\cite{GaussShape}}.}
	\label{dNdXi}
	\end{figure}

	The inclusive momentum distribution function of partons in jets $D(\xi)=\frac{dN}{d\xi}$ in NLLA is defined in terms of the variable $\xi=\ln(\frac{1}{x})$ where $x=\frac{p}{E_\mathit{jet}}$ and $p$ is the parton momentum. This distribution is predicted to have a distorted Gaussian shape {\cite{GaussShape}}:
	\begin{widetext}
	\begin{eqnarray}\label{eq:dXi}
	\frac{dN}{d\xi}=\frac{N}{\sigma\sqrt{2\pi}}\exp{\bigg[\frac{1}{8}l-\frac{1}{2}s\delta-\frac{1}{4}(2+l)\delta^{2}+\frac{1}{6}s\delta^{3}+\frac{1}{24}l\delta^{4}\bigg]},
	\end{eqnarray}
	\end{widetext}
 where $\delta=\xi-\xi_{0}$ and $\xi_{0}$ is the position of the maximum of the distribution. The coefficients $\sigma$, $s$, and $l$ are the width, skewness, and kurtosis of the inclusive momentum spectrum. These coefficients are calculated to next-to-leading order and depend on $Q_\mathit{eff}$. Overall, the function has three parameters to be determined experimentally: the normalization $N$, $Q_\mathit{eff}$, and an unknown higher-order correction term $O(1)$ {\cite{Webber}} in the calculation of $\xi_{0}=0.5\tau+a\sqrt{\tau}+O(1)$, where $a$ is a constant that depends on the number of colors and the number of effectively massless quarks. The unknown term $O(1)$ is expected to be independent of $\tau$. The predicted dependence of the inclusive momentum distribution on jet hardness is shown in Fig.~{\ref{dNdXi}}.

	The two-parton momentum correlation function $R(\xi_{1},\xi_{2})$ is defined to be the ratio of the two- and one-parton momentum distribution functions:
	\begin{eqnarray}\label{eq:r1}
	R(\xi_{1},\xi_{2})=\frac{D(\xi_{1},\xi_{2})}{D(\xi_{1})D(\xi_{2})},
	\end{eqnarray}
 	where 
	$D(\xi_{1},\xi_{2})=\frac{d^{2}N}{d\xi_{1}d\xi_{2}}$. The momentum distributions are normalized as follows: $\int D(\xi) d\xi=\left< n \right>$, where $\left< n \right>$ is the average multiplicity of partons in a jet, and $\int D(\xi_{1},\xi_{2})d\xi_{1}d\xi_{2}=\left< n(n-1) \right>$ for all pairs of partons in a jet. The average multiplicity of partons $\left< n \right>$ is a function of the dijet mass $M_{jj}$ and the size of the opening angle $\theta_{c}$. For $\theta_{c}=0.5$,  $\left< n \right>$ varies from $\sim6$ to $\sim12$ for $M_{jj}$ in the range 80--600 GeV/c$^{2}$ {\cite{SafonovPRL}}.

	 The Fong-Webber approximation of Eq.~(\ref{eq:r1}) for the two-parton momentum correlation function {\cite{Webber}} can be written as follows: 
	\begin{eqnarray}\label{eq:r2}
	R(\Delta\xi_{1},\Delta\xi_{2})=r_{0}+r_{1}(\Delta\xi_{1}+\Delta\xi_{2})+r_{2}(\Delta\xi_{1}-\Delta\xi_{2})^{2},
	\end{eqnarray} 
	where $\Delta\xi=\xi-\xi_{0}$, and the parameters $r_{0}$, $r_{1}$, and $r_{2}$ define the strength of the correlation and depend on the variable $\tau=\ln(Q/Q_\mathit{eff})$. Equation~(\ref{eq:r2}) is valid only for partons with $\xi$ around the peak of the inclusive parton momentum distribution, in the range $\Delta\xi \sim\pm 1$. The parameters $r_{0}$, $r_{1}$, and $r_{2}$ are calculated separately for quark and gluon jets from an expansion in powers of $1/\sqrt{\tau}$ using the assumption that the number of effectively massless quarks $N_f$ is 3. Keeping only terms controlled by theory, the parameters are:
	\begin{eqnarray}\label{eq:prq}
 r_{0}^{q}=1.75-\frac{0.64}{\sqrt{\tau}}, \ \ \ \ \   r_{1}^{q}=\frac{1.6}{\tau^{3/2}},  \ \ \ \ \ r_{2}^{q}=-\frac{2.25}{\tau^{2}},
	\end{eqnarray}
	\begin{eqnarray}\label{eq:prg}
 r_{0}^{g}=1.33-\frac{0.28}{\sqrt{\tau}}, \ \ \ \ \   r_{1}^{g}=\frac{0.7}{\tau^{3/2}},  \ \ \ \ \ r_{2}^{g}=-\frac{1.0}{\tau^{2}},
	\end{eqnarray}
	where $q$ and $g$ superscripts denote the correlation parameters for partons in quark jets and gluon jets, respectively.

	\begin{figure}
	\includegraphics[width=3.2in]
	{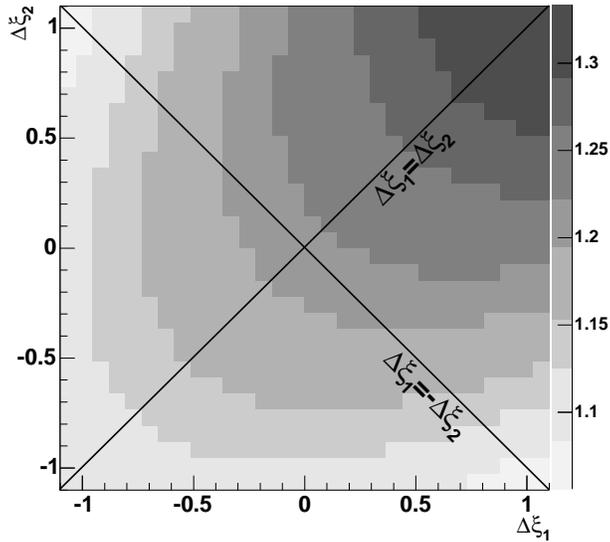}
	\caption{The NLLA parton momentum correlation function calculated for a gluon jet, $Q=50$ GeV, and $Q_\mathit{eff}=230$ MeV according to {\cite{Webber}}.}
	\label{CorrNLLA}
	\end{figure}

	The theoretical prediction of the shape of the two-parton momentum correlation distribution function is shown in Fig.~{\ref{CorrNLLA}}. Along the central diagonal $\Delta\xi_{1}=-\Delta\xi_{2}$, the shape of the two-parton momentum correlation is parabolic with a maximum at $\Delta\xi_{1}=\Delta\xi_{2}$. Along the central diagonal $\Delta\xi_{1}=\Delta\xi_{2}$, the shape is linear and increasing toward larger values of $\Delta\xi$, i.e. lower momentum partons. Therefore, the obvious features of the prediction are (1) the correlation should be stronger for partons with equal momenta $\Delta\xi_{1}=\Delta\xi_{2}$, and (2) the strength of this effect should increase for lower momentum partons.

	\subsection{Normalization}
	The correlation function from Eq.~(\ref{eq:r1}) includes two effects:  (1) multiplicity fluctuations of partons in a jet and (2) actual momentum correlations. In this analysis, we measure pure momentum correlations. This can be achieved by introducing one- and two-parton momentum distributions normalized to unity:
	\begin{eqnarray}
	D'(\xi)=\frac{D(\xi)}{\left<n\right>}, \ \ \ \ \  \int D'(\xi)d\xi=1,
	\end{eqnarray}
	\begin{eqnarray}
	D'(\xi_{1},\xi_{2})=\frac{D(\xi_{1},\xi_{2})}{\left<n(n-1)\right>}, \ \ \ \ \  \int D'(\xi_{1},\xi_{2})d\xi_{1}d\xi_{2}=1. 
	\end{eqnarray}
	
	Then, the correlation function can be defined as:
	\begin{widetext}
	\begin{eqnarray}\label{eq:C}
	C(\Delta\xi_{1},\Delta\xi_{2})=\frac{D'(\xi_{1},\xi_{2})}{D'(\xi_{1})D'(\xi_{2})}=\frac{\left<n\right>^{2}}{\left<n(n-1)\right>}R(\Delta\xi_{1},\Delta\xi_{2})=\frac{1}{F(\tau)}R(\Delta\xi_{1},\Delta\xi_{2}),
	\end{eqnarray} 
	\end{widetext}
	where $F(\tau)=\frac{\left<n(n-1)\right>}{\left<n\right>^{2}}$ is the second binomial moment. The explicit dependence of the binomial moments on the energy scale for quark and gluon jets is taken from theory {\cite{BinMoments}}:
	\begin{eqnarray}\label{eq:f}
	 F_{q}(\tau)=1.75-\frac{1.29}{\sqrt{\tau}}, \ \ \ \ \  F_{g}(\tau)=1.33-\frac{0.55}{\sqrt{\tau}}.
	\end{eqnarray}
	
	\subsection{Quark and gluon jets}
	In theory, correlation functions are calculated for quark and gluon jets separately and are denoted by $D_{q}(\xi)$ and $D_{g}(\xi)$, respectively. Since dijet events at the Tevatron consist of both quark and gluon jets, in order to compare data to theory, we rewrite the formula for the parton momentum distributions as follows:
	\begin{eqnarray}
	D(\xi)=f_{g}D_{g}(\xi)+(1-f_{g})D_{q}(\xi),
	\end{eqnarray} 
	\begin{eqnarray}
	D(\xi_{1},\xi_{2})=f_{g}D_{g}(\xi_{1},\xi_{2})+(1-f_{g})D_{q}(\xi_{1},\xi_{2}),
	\end{eqnarray} 
	where $f_{g}$ is a fraction of gluon jets in the sample. After simple algebraic transformations, it can be shown that the momentum correlation Eq.~(\ref{eq:C}) is reduced to the following:
	\begin{eqnarray}\label{eq:Cfinal}
	C(\Delta\xi_{1},\Delta\xi_{2})=c_{0}+c_{1}(\Delta\xi_{1}+\Delta\xi_{2})+c_{2}(\Delta\xi_{1}-\Delta\xi_{2})^{2},
	\end{eqnarray} 
	where the $c_{i}$ coefficients ($i=0,1,2$) are:
	\begin{eqnarray}\label{eq:c}
	c_{i}=\frac{f_{g}r^{2}}{f_{g}r^{2}F_{g}+(1-f_{g})F_{q}}r_{i}^{g}+\frac{1-f_{g}}{f_{g}r^{2}F_{g}+(1-f_{g})F_{q}}r_{i}^{q},
	\end{eqnarray} 
	where $r=\frac{\left<n_{g}\right>}{\left<n_{q}\right>}$ is the ratio of average multiplicities of partons in gluon and quark jets. The value of $r$ enters in the derivation of Eqs.~(\ref{eq:prq}) and (\ref{eq:prg}) \cite{Webber}, Eq.~(\ref{eq:f}) \cite{BinMoments}, and Eq.~(\ref{eq:c}). In NLLA this ratio is equal to 9/4. Values of $r$ by the {\sc pythia} 6.115 and {\sc herwig} 5.6 Monte Carlo generators are given in \cite{Sasha}. 

	\subsection{Local parton-hadron duality}	
	LPHD is a hadronization conjecture that suggests that the properties of hadrons and partons are closely related. In the simplest interpretation of LPHD, each parton at the end of the pQCD shower development picks up a color-matching partner from the vacuum sea and forms a hadron. Within LPHD, the momentum distributions of hadrons are related to those of partons via an energy-independent constant $K_{LPHD}$:
	\begin{eqnarray}
	\frac{dN_{hadrons}}{d\xi}=K_{LPHD}\cdot \frac{dN_{partons}}{d\xi}.
	\end{eqnarray}
	
	Past studies of inclusive particle distributions at $e^{+}e^{-}$ experiments  {\cite{Early}} and CDF {\cite{SafonovPRL,SafonovPRD}} have given strong support to the LPHD hypothesis. In this analysis, we extend the LPHD test by examining whether the two-particle momentum correlations predicted in the pQCD framework  also survive the hadronization. Note that in the two-particle momentum correlation given by Eq.~(\ref{eq:r1}), $K_{LPHD}$ simply cancels, suggesting that the correlation distributions for hadrons and partons are expected to be the same.

\section{CDF II detector}
Data used in this analysis were recorded with the CDF II detector. The detector was designed for precision measurements of the energy, momentum and position of particles produced in proton-antiproton collisions. This section provides a brief overview of the components relevant to our analysis. A detailed description of the entire detector can be found elsewhere {\cite{CDFII}}.

	CDF II uses a cylindrical coordinate system with the positive $z$ direction selected along the proton beam direction and azimuthal angle $\phi$ measured around the beam axis. The polar angle $\theta$  is measured with respect to the positive $z$ direction and the pseudorapidity $\eta$ is defined as $\eta=-\ln\left[\tan(\frac{\theta}{2})\right]$. 

	The CDF II tracking system is placed inside a 1.4 T solenoidal magnet. A Layer~00 single-sided silicon microstrip detector is mounted directly on the beam pipe, at an inner radius of $1.15$ cm and an outer radius of $2.1$ cm. A five-layer silicon microstrip detector (SVX II) is closest to the beamline, and is situated at a radial distance of 2.5 to 11 cm from the beam. The SVX II consists of three separate barrel modules with a combined length of 96 cm. Three of the five layers combine a $r$-$\phi$ measurement with a $z$-coordinate measurement while the remaining two layers combine $r$-$\phi$ with a small stereo angle of $1.2^{\circ}$. Three additional intermediate silicon layers (ISL) are positioned between 19 and 30 cm. The SVX II is surrounded by the central outer tracker (COT), an open-cell drift chamber providing up to 96 measurements of a charged particle track over the radial region from 40 to 137 cm. The 96 COT measurements come from 8 superlayers of 12 sense wires each. The superlayers alternate between axial and $3^{\circ}$ stereo. The pseudorapidity region covered by the COT is $\left|\eta\right|<1.0$. The momentum of a charged particle is determined by the curvature of its trajectory in the magnetic field.

	The CDF II tracking system is surrounded by calorimeters used to measure the energy of charged and neutral particles. In the central region the central electromagnetic (CEM), central hadronic (CHA), and wall hadronic calorimeters are made of lead (electromagnetic) and iron (hadronic) layers interspersed with scintillator. The pseudorapidity region covered by these calorimeters is $\left|\eta\right|<1.3$. The segmentation of the central calorimeters is $15^{\circ}$ in $\phi$ and 0.1 units in $\eta$. The measured energy resolutions for the CEM and CHA are $\sigma(E)/E=13.5\%/\sqrt{E_{T}}\oplus 2\%$ and $\sigma(E)/E=75\%/\sqrt{E_{T}}\oplus 3\%$, respectively. Here $E_{T}=E\sin{\theta}$ is transverse energy of an incident particle (electron for CEM and pion for CHA) and energies are measured in GeV.

\section{Event Selection}
\subsection{Triggers}
Events were collected using a single-tower trigger \cite{TrigTower} with a transverse energy $E_{T}$ threshold of 5 GeV (ST05) and with single jet triggers with $E_{T}$ thresholds of 20 (J020), 50 (J050), 70 (J070), and 100 (J100) GeV. Each of the jet triggers had a different sampling rate so as not to saturate the available trigger bandwidth.

\subsection{Jet reconstruction algorithm}
	Jets are reconstructed based on the calorimeter information using a cone algorithm {\cite{JetClu}}. The algorithm starts with the highest transverse energy tower and forms preclusters from an unbroken chain of continuous seed towers with transverse energy above 1 GeV within a window of $7\times7$ towers centered at the originating seed tower. If a seed tower is outside this window, it is used to form a new precluster. The coordinates of each precluster are the $E_{T}$-weighted sums of $\phi$ and $\eta$ of the seed towers within this precluster. In the next step, all towers with $E_{T}>0.1$ GeV within $R=\sqrt{(\Delta\phi)^{2}+(\Delta\eta)^{2}}=1.0$ of the precluster are merged into a cluster, and its $(\eta,\phi)$-coordinates are recalculated. This procedure of calculating cluster coordinates is iterated until a stable set of clusters is obtained. A cluster is stable when the tower list is unchanged from one iteration to the next. If the clusters have some finite overlap, then an overlap fraction is computed as the sum of the $E_T$ of the common towers divided by the $E_T$ of the smaller cluster. If the fraction is above a cutoff value equal to 0.75, then the two clusters are combined. If the fraction is less than the cutoff, the shared towers are assigned to the closest cluster. The raw energy of a jet is the sum of the energies of the towers belonging to the corresponding cluster. Corrections are applied to the raw energy to compensate for the non-linearity and non-uniformity of the energy response of the calorimeter, the energy deposited inside the jet cone from sources other than the leading parton, and the leading parton energy deposited outside the jet cone. A detailed description of this procedure can be found in {\cite{JetCluCorr}}. 

\subsection{Offline selection}
 	Cosmic ray events are rejected by applying a cutoff on the missing transverse energy \met significance {\cite{missingEt}}, defined as \met$/\sqrt{\Sigma E_{T}}$, where $\Sigma E_{T}=\Sigma_{i} E_{T}^{i}$ is the total transverse energy of the event, as measured using calorimeter towers with $E_{T}^{i}$ above $100$ MeV. The thresholds are 3.0 GeV$^{1/2}$ for data collected using a single tower trigger with $E_{T}$ threshold of 5 GeV, and  3.5, 5.0, 6.0, and 7.0 GeV$^{1/2}$  for data collected using jet triggers with thresholds of 20, 50, 70, and 100 GeV, respectively.

To ensure fully efficient vertex and track reconstruction, we require only one vertex in the event with $\left|z\right|<60$ cm. 
	\begin{figure}
	\includegraphics[width=3.5in]
	{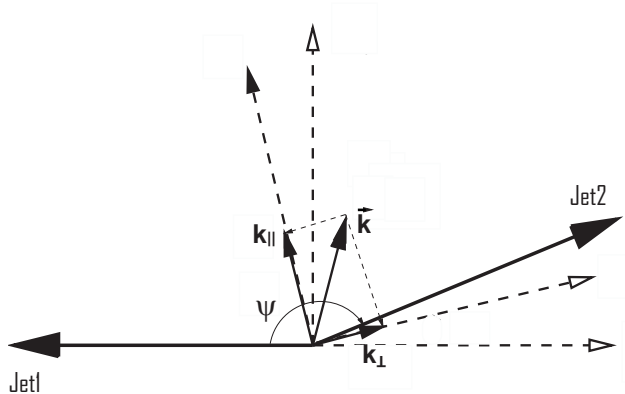}
	\caption{ Definition of variables for the jet balance requirement. The plane perpendicular to the beamline is shown. The vector $\vec{k}$ represents a vector sum of the two leading jets' momenta. The $k_{||}$ and $k_{\perp}$ components of $\vec{k}$ are parallel and perpendicular to the bisector of the two jets, respectively.}
	\label{BalCut}
	\end{figure}

	To ensure robust and high efficiency track reconstruction and applicability of the background removal technique (see Sec.~V(B)), only events with both leading jets in the central region ($\left|\eta\right|<0.9$) are selected. To reject events with poorly measured jets, we require the two leading jets to be well balanced in $E_{T}$: $k_{\perp}/ (E_{T}^{1}+E_{T}^{2}) < 2\sigma_{k_{\perp}}$. Here $E_{T}^{1}$ and $E_{T}^{2}$ are the transverse energies of the first and the second leading jets, respectively, and $k_{\perp}$ is:
	\begin{eqnarray}
	k_{\perp}=\sqrt{k^{2}-k_{||}^{2}},
	\end{eqnarray} 
	\begin{eqnarray}
	k_{||}=(E_{T}^{1}+E_{T}^{2})\cdot \cos(\psi/2),
	\end{eqnarray} 
where $\vec{k}$ is a vector sum of momenta of the two leading jets, $\psi$ is the angle between the two leading jets, and $\sigma_{k_{\perp}}$ is the resolution of $k_{\perp}$. The definitions of $\vec{k}$, $k_{\perp}$, and $k_{||}$ are illustrated in Fig.~{\ref{BalCut}}. The component $k_{\perp}$ is known to be sensitive to jet energy mismeasurements, while $k_{||}$ is more sensitive to hard gluon radiation.

	In events with high energy jets, a single particle emerging from a jet at a sufficiently large angle with respect to the jet axis can be mistakenly identified as a separate jet. A jet can also be produced from the underlying event. Therefore, rejection of all events with more than two jets can introduce possible biases. We allow up to two extra jets, but their energy is required to be small: $E_{T}^\mathit{extra}<5.5$ GeV $+0.065(E_{T}^{1}+E_{T}^{2})$, where $E_{T}^\mathit{extra}$ is the transverse energy of an extra jet.

	After application of the event selection cuts, the final sample consists of approximately 250,000 events and is further divided into seven bins according to the dijet mass as measured by the calorimeters and defined as:
\begin{eqnarray}
M_{jj}=\sqrt{(E_{1}+E_{2})^{2}/c^{4}-(\vec{P}_{1}+\vec{P}_{2})^{2}/c^{2}},
\end{eqnarray}
 where $E$ and $\vec{P}$ are the energies and momenta of the two leading jets, respectively.

	\begin{table*}
	\caption{ Dijet mass bins boundaries, average dijet invariant mass $\left<M_{jj}\right>$, average $E_{jet}$-weighted jet hardness Q, and number of events in each bin after the event selection requirements N$_\mathit{ev}$.}
	\begin{ruledtabular}
	\begin{tabular}{ccccccc}
	Bin  & Trigger & Low edge (GeV/$c^2$) & High edge (GeV/$c^2$) & 
	{$\left<M_{jj}\right>$ (GeV/$c^2$)} & Q (GeV) & N$_\mathit{ev}$\\ \hline
	1 & ST05 & 66 & 95  & 76  & 19 & 15229\\
	2 & J020 & 95 & 132 & 108 & 27 & 77246\\
	3 & J020 & 132& 180 & 149 & 37 & 17682\\
	4 & J050 & 180& 243 & 202 & 50 & 80608\\
	5 & J050 & 243& 323 & 272 & 68 & 18528\\
	6 & J070 & 323& 428 & 361 & 90 & 12000\\
	7 & J100 & 428& 563 & 475 & 119& 19150\\
	\end{tabular}
	\end{ruledtabular}
	\label{massbindef}
	\end{table*}
	 The mass bin boundaries, average invariant mass $\left<M_{jj}\right>$, and number of events in each bin are given in Table~{\ref{massbindef}}. The bin width is selected to be $3\cdot \sigma_{M_{jj}}$, where $\sigma_{M_{jj}}$ is the calorimeter resolution for the dijet mass determination, $\frac{\sigma_{M_{jj}}}{M_{jj}}\sim$10--15\%. Events with dijet invariant mass below $66$ GeV/$c^2$ are not used in the measurement because of the lower trigger efficiency.

\subsection{Systematic uncertainties associated with the event selection}
   The sensitivity of the two-particle momentum correlation parameters $c_{0}$, $c_{1}$, and $c_{2}$ to various uncertainties in the event selection procedure is evaluated as follows. For each source of systematic uncertainty the so-called ``default'' and ``deviated'' two-particle momentum correlation distributions are obtained. The default distribution is produced using the standard selection requirements described in this article. Then, the deviated distribution is obtained by varying all relevant parameters according to the estimated systematic uncertainty (one source of uncertainty at a time). For each bin in correlation $C(\Delta\xi_{1},\Delta\xi_{2})$, a scale factor is calculated by taking the bin-by-bin ratio of the deviated and default distributions:

	\begin{eqnarray}
	\epsilon=\frac {C(\Delta\xi_{1},\Delta\xi_{2})_{deviated}}{C(\Delta\xi_{1},\Delta\xi_{2})_{default}}.
	\end{eqnarray}

  	The difference between correlation distributions in the data with and without this bin-by-bin scale factor is taken as a measure of the systematic uncertainty:

	\begin{eqnarray}
	\Delta C(\Delta\xi_{1},\Delta\xi_{2})_{Data}=\left|(1-\epsilon) \cdot C(\Delta\xi_{1},\Delta\xi_{2})_{Data} \right|. 
	\end{eqnarray}

	Further in this section, we discuss sources of systematic uncertainties at the level of the event selection; their contributions to the values of $c_{0}$, $c_{1}$, and $c_{2}$ are given in Table {\ref{System}}.

	\begin{table}
	\caption{Summary of the systematic uncertainties of the correlation parameters $c_{0}$, $c_{1}$, and $c_{2}$ for the dijet mass bin with $Q=50$ GeV.}
	\begin{ruledtabular}
	\begin{tabular}{lccc}
	Origin of systematic uncertainty  &  $\Delta c_{0}$  & $\Delta c_{1}$   &  $\Delta c_{2}$
	\\ \hline
	Luminosity dependence & 0.001 & 0.004 & 0.002\\
	Jet energy scale & 0.001 & 0.001 & 0.001\\
	Balance and extra jet cuts & 0.006 & 0.001 & 0.003\\
	Mismeas. of jet direction & 0.006 & 0.008 & 0.007\\
	Track quality cuts & 0.014 & 0.008 & 0.006\\
	Underlying event background & 0.001 & 0.004 & 0.001\\
	Tracking inefficiency & 0.011 & 0.001 & 0.002\\
	Neutral particles & 0.002 & 0.002 & 0.001\\
	\end{tabular}
	\end{ruledtabular}
	\label{System}
	\end{table}

	In each trigger sample only the events with trigger efficiency higher than 99$\%$ are used. To check that trigger effects do not bias the measurement, we verify the continuity of the distributions of particle multiplicity in a jet in the transition between adjacent dijet trigger samples. No detectable offsets are observed.

	To evaluate the uncertainty due to the value of the parameter $R$ of the jet reconstruction algorithm, we compare the results of the measurement using three different values of $R$ (0.4, 0.7, 1.0). This effect proved to be small compared to the other sources of systematic uncertainty.

We require only one vertex in the event, which greatly reduces the contribution of multiple proton-antiproton interactions in the same bunch crossing. However, in some cases two vertices can be very close to each other and be reconstructed as a single vertex. This can become significant at high instantaneous luminosity. To evaluate the uncertainty due to this effect, we divide each dijet mass bin into smaller bins based on the instantaneous luminosity. Momentum correlation distributions are compared in these smaller samples and the difference is taken as a measure of the systematic uncertainty.

To evaluate the uncertainty due to the jet energy corrections we use parameterizations in which the jet energy scale is shifted by plus or minus one standard deviation {\cite{JetCluCorr}}. We then reclassify the events according to their dijet mass. The difference between the default and deviated distribution is assigned to be the systematic uncertainty.

We use Monte Carlo dijet samples produced by {\sc pythia} tune A to study systematic uncertainties associated with the jet balance requirement, the number of allowed extra jets, and their energy. The default two-particle momentum correlation distribution is compared to the deviated one. The deviated distribution has no requirements imposed on the jet balance or on the extra jet number and extra jet energy.

Monte Carlo simulations are used to evaluate the systematic uncertainty due to mismeasurement of the jet direction. Two-particle momentum correlations are compared for two cases. In one case particles are counted in a restricted cone around the jet direction as determined by the detector response in the simulation. In the second case the direction of primary partons from the hard scattering as given by {\sc pythia} tune A is used for the cone axis.

 	\section{Track Selection, Corrections, Systematic Uncertainties}
	Measurements described below are performed in the dijet center of mass frame. For Lorentz boosts all particles are treated as pions. Experimentally we define the variable $\xi$ as $\xi=\ln (1/x)=\ln \frac{E_{jet}}{p_{track}}$, where $E_{jet}$ is the jet energy as measured by the calorimeters and $p_{track}$ is the track momentum as measured by the tracking system. The correlation distributions are measured for all track pairs that pass track quality requirements and lie within a restricted cone of opening angle $\theta_{c}=0.5$ radians relative to the jet axis. The peak position of the inclusive momentum distribution $\xi_{0}$ is constant for a given jet hardness $Q$ and is obtained from the data. The measurements are corrected for various backgrounds both correlated and uncorrelated with jet direction.
 	
	\subsection{Track quality requirements}
	Several selection requirements are applied to ensure that the tracks in the measurement originate at the primary vertex and are not produced by cosmic rays, multiple $p\bar p$ interactions within the same bunch crossing, $\gamma$-conversions, $K^{0}$ and $\Lambda$-decays, or other types of backgrounds.

	In our analysis we require full three-dimensional track reconstruction. The description of CDF II track reconstruction can be found in {\cite{CDFIItdr,ionim}}. Poorly reconstructed and spurious tracks are removed by requiring a good track fitting parameter $\chi^{2}_{COT}<6.0$. Charged particles are required to have transverse momentum $p_{T}>0.3$ GeV/c.

	The parameter $\Delta z$ is defined as the difference between the $z$ position of the track at the point of its closest approach to the beamline and the $z$ position of the primary vertex. This parameter is used to remove tracks not originating at the primary interaction by requiring $\left| \Delta z \right|<5 \cdot \sigma_{\Delta z}$, where $\sigma_{\Delta z}$ is determined for different categories of tracks based on the number of SVX II, ISL, and COT hits.

\begin{figure}
	\includegraphics[width=2.0in]
	{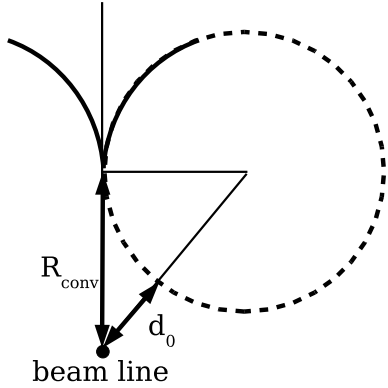}
	\caption{Schematic illustration of the distance $R_{conv}$ from the beamline to the point where the conversion occurred. Here, $d_{0}$ is the impact parameter.}
	\label{Rconv}
	\end{figure}

	Tracks produced from $\gamma$-conversions are removed using a combination of requirements on impact parameter $d_{0}$ and the distance $R_{conv}$ (see Fig.~{\ref{Rconv}}). The impact parameter $d_{0}$ is defined as the shortest distance in the $r-\phi$ plane between the beamline and the trajectory of the particle obtained by the tracking algorithm fit. It can be shown that for electrons and positrons originating from $\gamma$-conversion:
	\begin{eqnarray}
	R_{conv} \approx \sqrt{\frac{d_{0}p_{T}}{0.15B}},
	\end{eqnarray}
where $p_{T}$ is the transverse momentum of the charged particle in GeV/c, $B$ is the magnetic field in Tesla and $R_{conv}$ is measured in meters. Monte Carlo studies indicate that the combined requirement of $\left|d_{0}\right|<5\cdot\sigma_{d_{0}}$ or $R_{conv}<0.13$ m is more efficient at removing $\gamma$-conversion tracks than the $d_{0}$ requirement alone. The resolution of the impact parameter, $\sigma_{d_{0}}$, is evaluated for different categories of tracks based on the number of SVX II, ISL, and COT hits. The value $R_{conv}=0.13$ m is motivated by the location of SVX II readout electronics. Indeed, conversions occurring at this radius are clearly seen in the data. 

	\begin{figure}
	\includegraphics[width=4.2in]
	{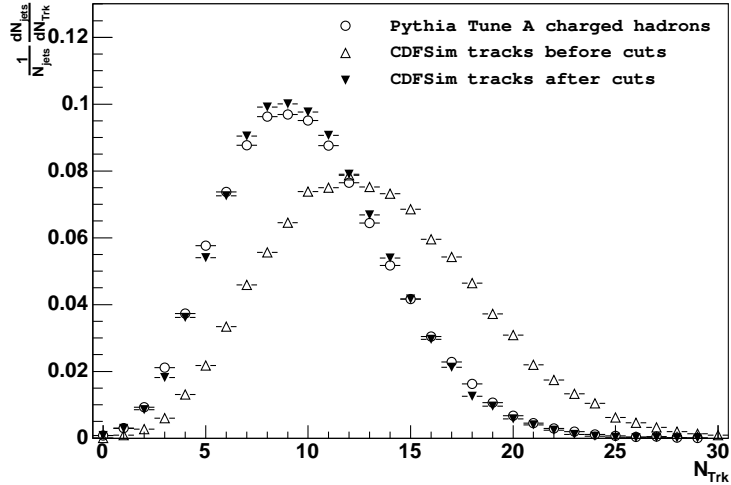}
	\caption{Monte Carlo track multiplicity in jets before and after applying track quality cuts. The distributions are for the dijet mass bin with $Q=50$ GeV. Particles are counted within a cone of opening angle $\theta_{c}=0.5$ radians. {\sc cdfsim} refers to the full CDF II detector simulation.}
	\label{Nave}
	\end{figure}

	\begin{figure}
	\includegraphics[width=4.2in]
	{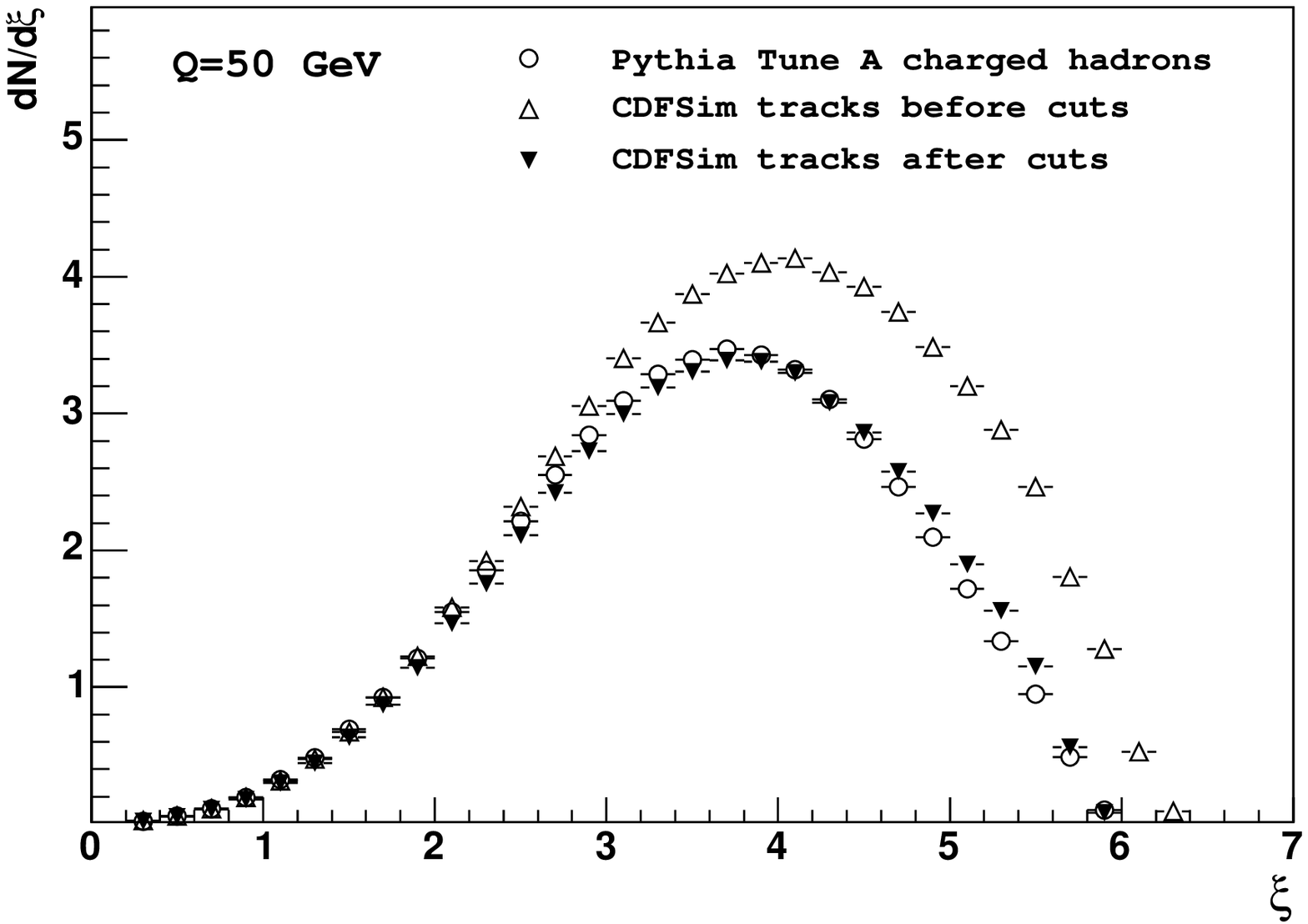}
	\caption{Inclusive momentum distributions of Monte Carlo tracks in jets before and after applying track quality cuts. The distributions are for the dijet mass bin with $Q=50$ GeV. Particles are counted within a cone of opening angle $\theta_{c}=0.5$ radians. {\sc cdfsim} refers to the full CDF II detector simulation.}
	\label{Xiave}
	\end{figure}

	To verify the effectiveness of the track quality cuts, we compare distributions of the inclusive particle multiplicity and momentum in {\sc pythia} tune A at the generator level and at the level of the detector simulation ({\sc cdfsim}). The comparison is shown in Figs.~{\ref{Nave}} and {\ref{Xiave}}. {\sc cdfsim} propagates particles through the detector including both conversions and in-flight decays to simulate the CDF II detector response. The agreement after selection cuts are applied confirms that the cuts do remove most of the background tracks. The effect of the remaining fraction of secondary tracks is estimated by comparing the correlation distributions $C(\Delta\xi_{1},\Delta\xi_{2})$ at the charged hadron level and the {\sc cdfsim} level and producing a corresponding bin-by-bin scale factor. The difference between distributions in data with and without this scale factor applied is assigned as the systematic uncertainty associated with the track quality cuts.
	\subsection{Underlying event background subtraction}
	Generally, tracks from the underlying event tend to dilute the two-particle momentum correlation. It is not possible to correct for this effect on an event-by-event basis, but the average correction factor can be reconstructed statistically. In order to correct for the underlying event contribution, we apply the following procedure. On an event-by-event basis, two complementary cones are positioned at the same polar angle with respect to the beamline as the original dijet axis but in the plane perpendicular to the dijet axis as shown in Fig.~{\ref{ComplCone}. Complementary cones defined this way are at $90^{\circ}$ in $\phi$ (i.e. as far as possible) from the dijet axis. This can be done when the dijet axis is within $45^{\circ}<\theta <135^{\circ}$, and this condition is automatically satisfied by our event selection. We assume that cones formed in such a fashion collect statistically the same amount of background (which is uncorrelated with jets) as the cones around the jet axis {\cite{SafonovPRD}}. 
	
	\begin{figure}
	\includegraphics[width=3.7in]
	{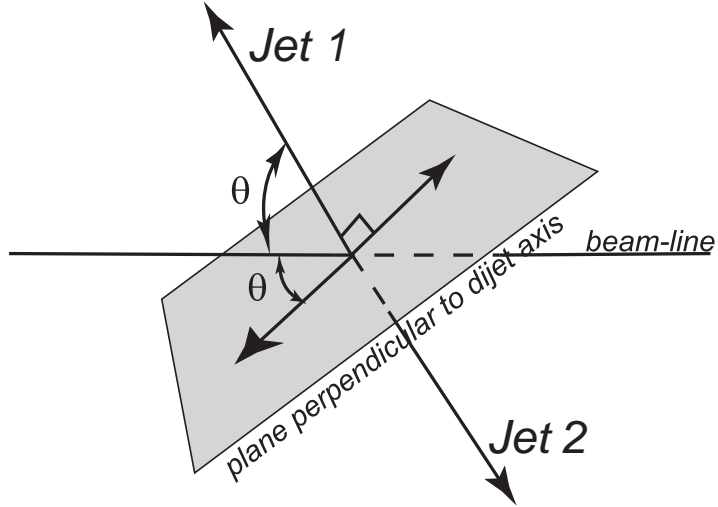}
	\caption{Illustration of the definition of complementary cones. The unlabeled arrows are the axes of the cones complementary to jets 1 and 2. The complementary cone makes the same angle $\theta$ with the beamline as the jet cone.}
	\label{ComplCone}
	\end{figure}

   In order to obtain the corrected expression for $C(\Delta\xi_{1},\Delta\xi_{2})$, one needs to subtract the background from the one- and two-particle momentum distributions. This can be achieved by considering particles in jet cones together with particles in complementary cones. It can be shown that the momentum distributions after background subtraction $\tilde{D}$ are:
	\begin{widetext}
	\begin{eqnarray}
	\tilde{D}(\xi)=D_{jet}(\xi)-D_{compl}(\xi),
	\end{eqnarray}
	\begin{eqnarray}
	\tilde{D}(\xi_{1},\xi_{2}) \approx 2D_{jet}(\xi_{1},\xi_{2})-D_{jet+compl}(\xi_{1},\xi_{2})+2D_{compl}(\xi_{1},\xi_{2}),
	\end{eqnarray}
	\end{widetext}
	where the $jet$ subscript denotes the distribution for particles in jet cones, $compl$ denotes the distribution for particles in complementary cones, and $jet+compl$ denotes the distribution for the combined set of particles in either jet cones or complementary cones.

	To evaluate systematic uncertainties associated with the background subtraction using the complementary cones, we use the following procedure. The amount of background in a jet cone is increased by a factor of two by adding tracks from the complementary cone of another event. Then, the background subtraction procedure described above is applied taking into account the artificially doubled background. After the subtraction the correlation distribution is expected to be the same as the distribution using the original background. The difference between the two-particle momentum correlation distributions obtained after the subtraction of either the original or the doubled background is assigned as a measure of the systematic uncertainty.

 	\subsection{Tracking inefficiency}
	A high efficiency of track reconstruction is ensured by selecting events with central jets. However, there still may be non-reconstructed tracks inside the jet. To evaluate the corresponding systematic uncertainty, we have modeled the track reconstruction inefficiency using the function $P(\xi)=p_{1}+p_{2}\xi$, which denotes the probability of losing a track with given $\xi$. Values of the parameters $p_{1}$ and $p_{2}$ were varied over a range far exceeding the estimated COT inefficiency. The correlation distributions show a very weak dependence on tracking inefficiency. The range of momentum correlation variation in this tracking inefficiency model is taken as a measure of the systematic uncertainty (see Table {\ref{System}}).

	\subsection{Neutral particles}
	Theoretical predictions of correlation distributions are done at the parton level, while LPHD relates final partons to hadrons, assuming that all hadrons are counted. The analysis, however, is done for charged particles only. To estimate the effect of neutral particles the momentum correlation in a {\sc pythia} tune A sample is compared for charged particles and all particles. The difference is assigned as the corresponding systematic uncertainty (see Table {\ref{System}}).	

	\subsection{Resonance decays}
	The presence of resonance decays may be expected to cause differences between the correlation in data and the theoretical predictions. We examine this effect by comparing the correlations in Monte Carlo events for hadrons before and after resonance decays. We find that this results in insignificant changes in $C(\Delta\xi_{1},\Delta\xi_{2})$ and does not change the overall level of the correlation.

	\subsection{Heavy flavor jets}
	Theoretical predictions of correlation distributions are obtained for jets originated from gluons or light quarks only. In the data sample we expect a small fraction ($\sim5$$\%$) of heavy flavor jets. To estimate the size of this effect we repeat the analysis with the assumption that the correlations in heavy flavor jets are same as in gluon jets. This translates into a $3$ MeV change in the value of $Q_\mathit{eff}$ and is negligibly small compared to the size of the systematic uncertainty.

\section{NLLA fits to data}
	The inclusive momentum distributions $D(\xi)=\frac{dN}{d\xi}$ in all seven experimental dijet mass bins are simultaneously fit to the theoretical Fong-Webber function. In the fit the $Q_\mathit{eff}$ and $O(1)$ parameters are required to have same value in all dijet mass bins while the normalization parameter $N(\tau)$ is allowed to vary from one bin to another. Figure~{\ref{MomDataNLLA}} shows the distributions in data corresponding to the dijet mass bins with $Q=27$, $50$, and $90$ GeV. The error bars correspond to both the statistical and systematic uncertainties added in quadrature. The solid curves correspond to the fit of the data to the theoretical Fong-Webber function, and the dashed curves represent the extrapolations out of the fit regions. The extracted values of the fit parameters are $Q_\mathit{eff}=180\pm 40$~MeV and $O(1)=-0.6\pm 0.1$, where the uncertainties are statistical and systematic added in quadrature. The value of $Q_\mathit{eff}$ is consistent with the results of a previous CDF measurement \cite{SafonovPRD}.

	The two-particle momentum correlation distributions $C(\Delta\xi_{1},\Delta\xi_{2})$ are produced for seven bins of dijet mass and do show the shape predicted by theory. In this paper we plot the central diagonal profiles $\Delta\xi_{1}=-\Delta\xi_{2}$ and $\Delta\xi_{1}=\Delta\xi_{2}$ (see Fig.~{\ref{CorrNLLA}}) of the distributions. Figures~{\ref{CorrDataNLLA1}}, {\ref{CorrDataNLLA4}}, and {\ref{CorrDataNLLA7}} show the distributions corresponding to the dijet mass bins with $Q=27$, $50$, and $90$ GeV, respectively. The bin size $\Delta\xi=0.2$ is chosen to be much wider than the momentum resolution in the fitted range. The smaller error bars correspond to the statistical uncertainty only, while the larger error bars correspond to both the statistical and systematic uncertainties added in quadrature. The 2-dimensional momentum correlation distribution is fit according to Eq.~(\ref{eq:Cfinal}) with three free parameters $c_{0}$, $c_{1}$, and $c_{2}$. The solid lines in Figs.~{\ref{CorrDataNLLA1}}, {\ref{CorrDataNLLA4}}, and {\ref{CorrDataNLLA7}} show the profiles of the fit functions. The extracted values of the fit parameters are given in Table {\ref{Csummary}}. The fit range $-1<\Delta\xi<1$ is motivated by the region of validity of the NLLA calculations.

	The dash-dotted lines in Figs.~{\ref{CorrDataNLLA1}}, {\ref{CorrDataNLLA4}}, and {\ref{CorrDataNLLA7}} correspond to the theoretical curves given by Eq.~(\ref{eq:Cfinal}) for $Q_\mathit{eff}=180\pm 40$~MeV, extracted from fits of the inclusive momentum distributions. The dashed lines correspond to the results of the Perez-Ramos calculation  for the value of $Q_\mathit{eff}=230\pm 40$~MeV extracted from fits of the inclusive momentum distributions to the MLLA function \cite{SafonovPRD}. The fraction of gluon jets in the sample, used to model the theoretical prediction for quark and gluon jets, is obtained using {\sc pythia} tune A with CTEQ5L parton distribution functions {\cite{CTEQ5L}}.

	The systematic uncertainty due to the parton distribution functions is evaluated by comparing results for the fraction of gluon jets $f_{g}$ obtained using CTEQ5L and CTEQ6.1 \cite{CTEQ61} PDF sets. The systematic uncertainty due to the value of $r$ was evaluated by taking the difference between the theoretical value ($r_{theory}=9/4$), used as default, and experimental value ($r_{exp}=1.8$) {\cite{Sasha}} and propagating it to the value of $Q_\mathit{eff}$. Both systematic uncertainties were found to be negligible.

	The overall qualitative agreement between the data and the Fong-Webber calculation \cite{Webber} is very good. The data follow the theoretical trends and show an enhanced probability of finding two particles with the same value of momenta (indicated by the parabolic shape of the $\Delta\xi_{1}=-\Delta\xi_{2}$ central diagonal profile with its maximum at $\Delta\xi_{1}=\Delta\xi_{2}=0$). This effect becomes larger for particles with lower momenta (the positive slope of the $\Delta\xi_{1}=\Delta\xi_{2}$ central diagonal profile). An offset in the overall level of correlation is observed in all seven dijet mass bins, indicating that the Fong-Webber prediction overestimates the parameter $c_{0}$ of the correlation. The Perez-Ramos curves \cite{Redamy} qualitatively show the same trends; however, the quantitative disagreement is obviously larger for the Perez-Ramos predictions compared to the Fong-Webber predictions \cite{Webber}. 
	\begin{figure}
	\includegraphics[width=2.7in]
	{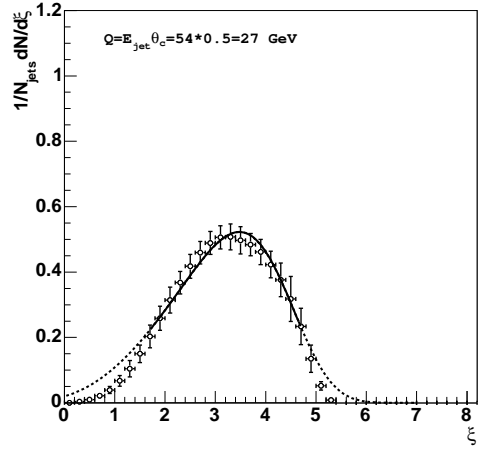}
	\\
	\includegraphics[width=2.7in]
	{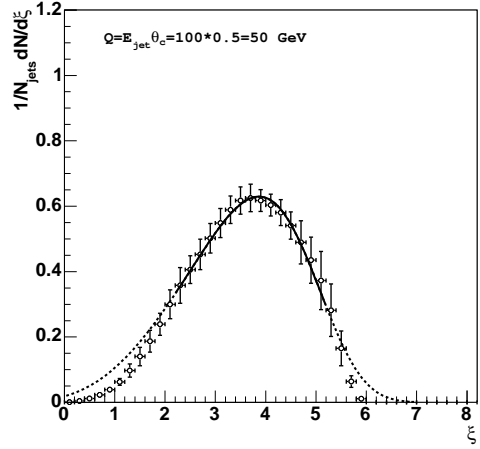}
	\\
	\includegraphics[width=2.7in]
	{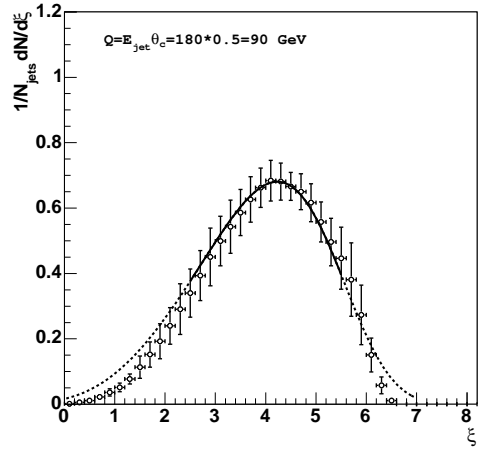}
	\caption{ Inclusive momentum distributions of particles in jets in the restricted cone of size $\theta_{c}=0.5$ radians for the dijet mass bin with $Q=27$ GeV (top), $Q=50$ GeV (middle), and $Q=90$ GeV (bottom). The solid curves correspond to the fit of CDF data to the theoretical Fong-Webber function (as calculated in {\cite{Webber}}), the dashed curves represent the extrapolations out of the fit regions.}
	\label{MomDataNLLA}
	\end{figure}

	\begin{table*}
	\caption{Summary of the correlation parameters $c_{0}$, $c_{1}$, and $c_{2}$ measured in seven dijet mass bins. The first uncertainty is statistical, the second one is systematic.}
	\begin{ruledtabular}
	\begin{tabular}{cccc}
	Q (GeV)  &  $c_{0}$  &   $c_{1}$  &   $c_{2}$  
	\\ \hline
	19  &  $1.078\pm0.007\pm0.016$ & $0.081\pm0.006\pm0.016$ & $-0.047\pm0.006\pm0.008$\\
	27  &  $1.076\pm0.003\pm0.022$ & $0.068\pm0.002\pm0.015$ & $-0.038\pm0.002\pm0.012$\\
	37  &  $1.075\pm0.005\pm0.018$ & $0.057\pm0.004\pm0.013$ & $-0.031\pm0.004\pm0.012$\\
	50  &  $1.079\pm0.002\pm0.019$ & $0.051\pm0.002\pm0.014$ & $-0.029\pm0.002\pm0.010$\\
	68  &  $1.081\pm0.004\pm0.028$ & $0.040\pm0.004\pm0.012$ & $-0.027\pm0.004\pm0.011$\\
	90  &  $1.081\pm0.005\pm0.023$ & $0.046\pm0.004\pm0.015$ & $-0.024\pm0.004\pm0.014$\\
	119 &  $1.077\pm0.004\pm0.033$ & $0.028\pm0.003\pm0.013$ & $-0.019\pm0.003\pm0.015$\\
	\end{tabular}
	\end{ruledtabular}
	\label{Csummary}
	\end{table*}

	\begin{figure}
	\includegraphics[width=2.8in]
	{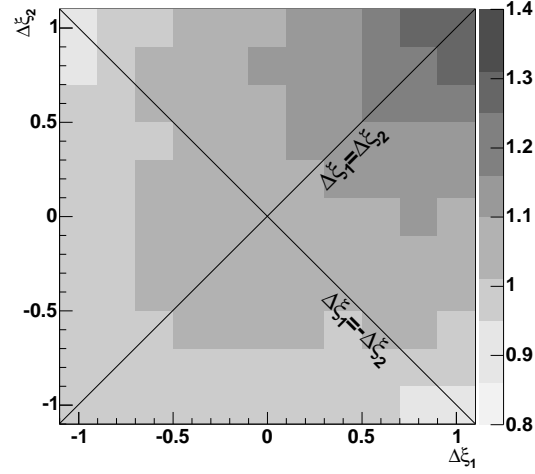}
	\\
	\includegraphics[width=2.8in]
	{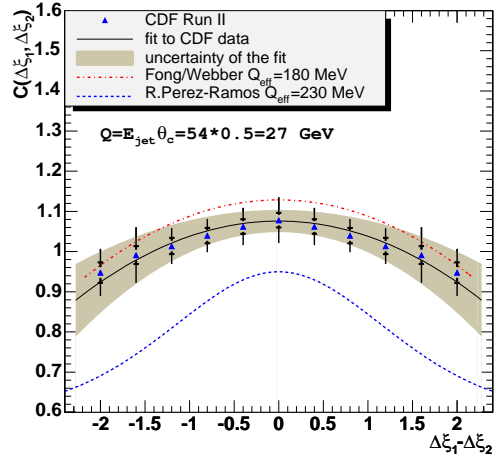}
	\\
	\includegraphics[width=2.8in]
	{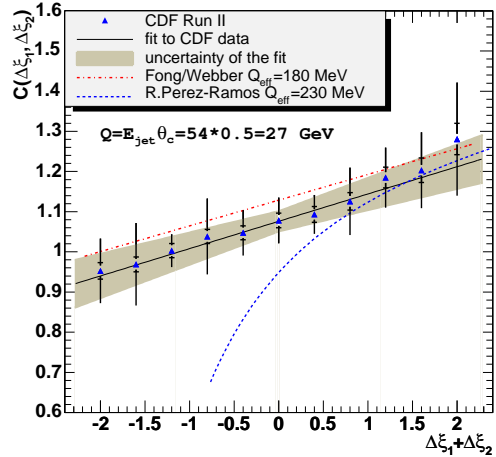}
	\caption{ Two-particle momentum correlations in jets in the restricted cone of size $\theta_{c}=0.5$ radians for the dijet mass bin with $Q=27$ GeV (top). Central diagonal profiles $\Delta\xi_{1}=-\Delta\xi_{2}$ (middle) and $\Delta\xi_{1}=\Delta\xi_{2}$ (bottom) of the distributions are shown. The correlation in data is compared to that of theory (as calculated in  {\cite{Webber}} for $Q_\mathit{eff}=180$~MeV and in {\cite{Redamy}} for $Q_\mathit{eff}=230$~MeV).}
	\label{CorrDataNLLA1}
	\end{figure}
	
	\begin{figure}
	\includegraphics[width=2.8in]
	{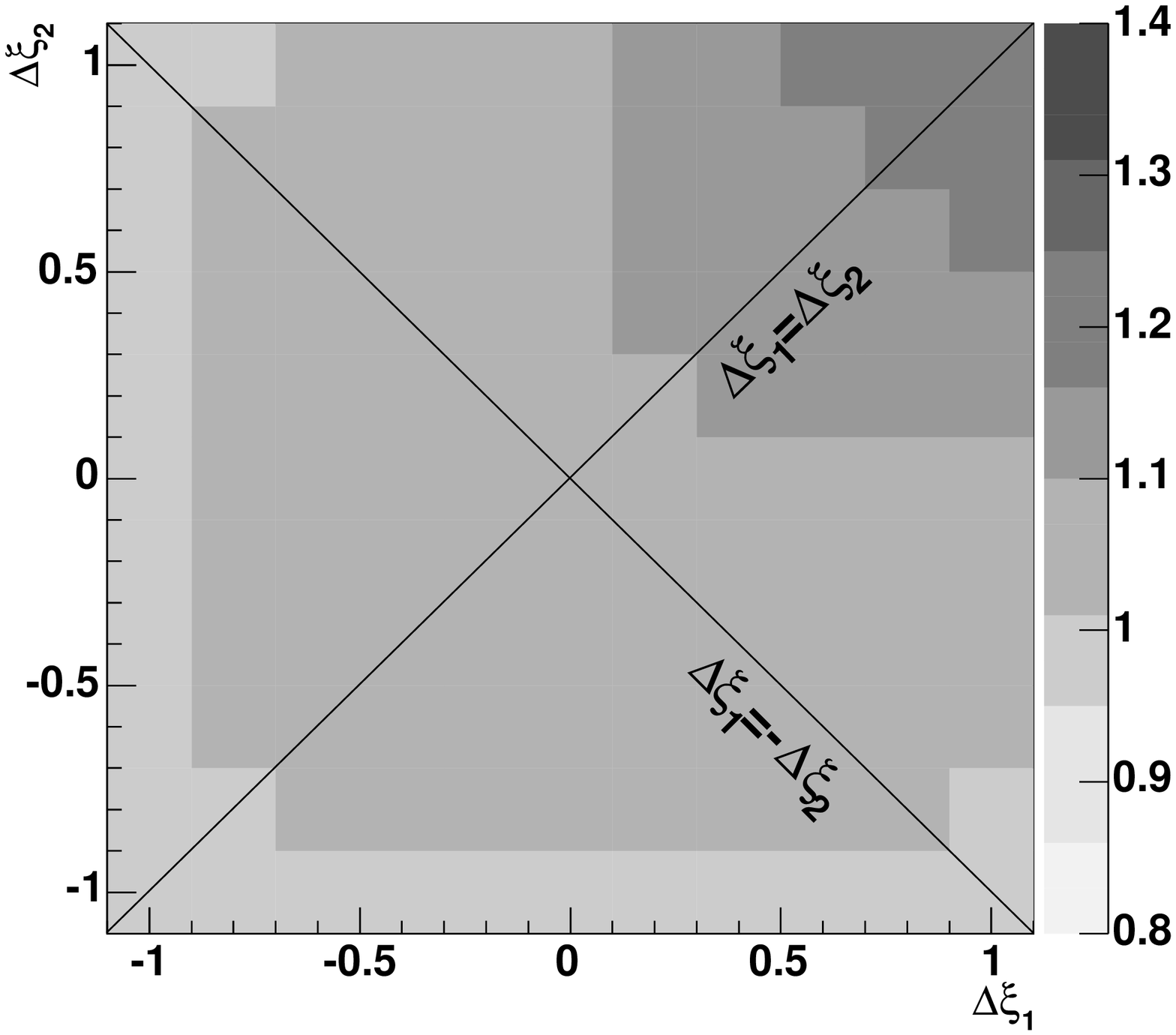}
	\\
	\includegraphics[width=2.8in]
	{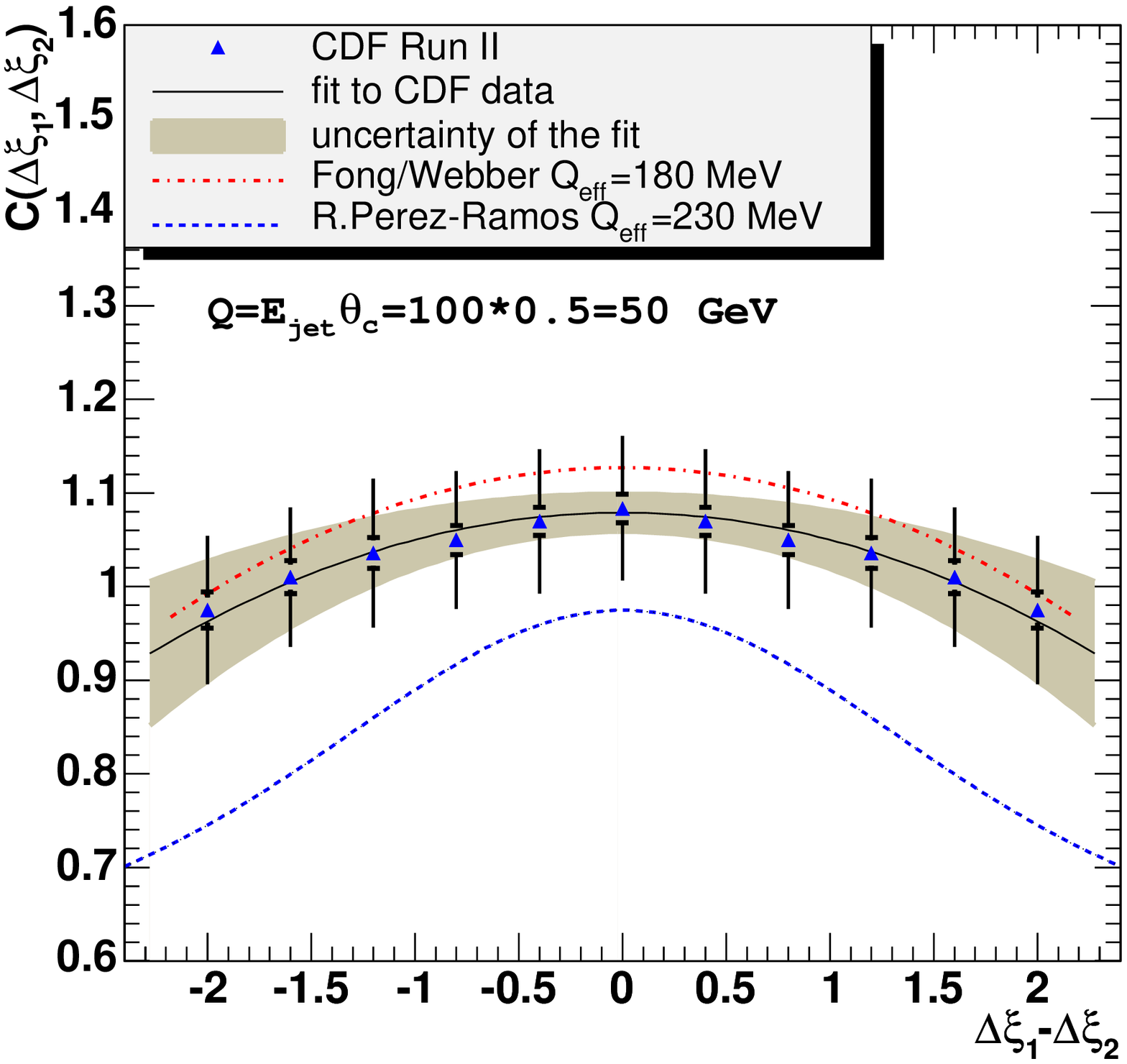}
	\\
	\includegraphics[width=2.8in]
	{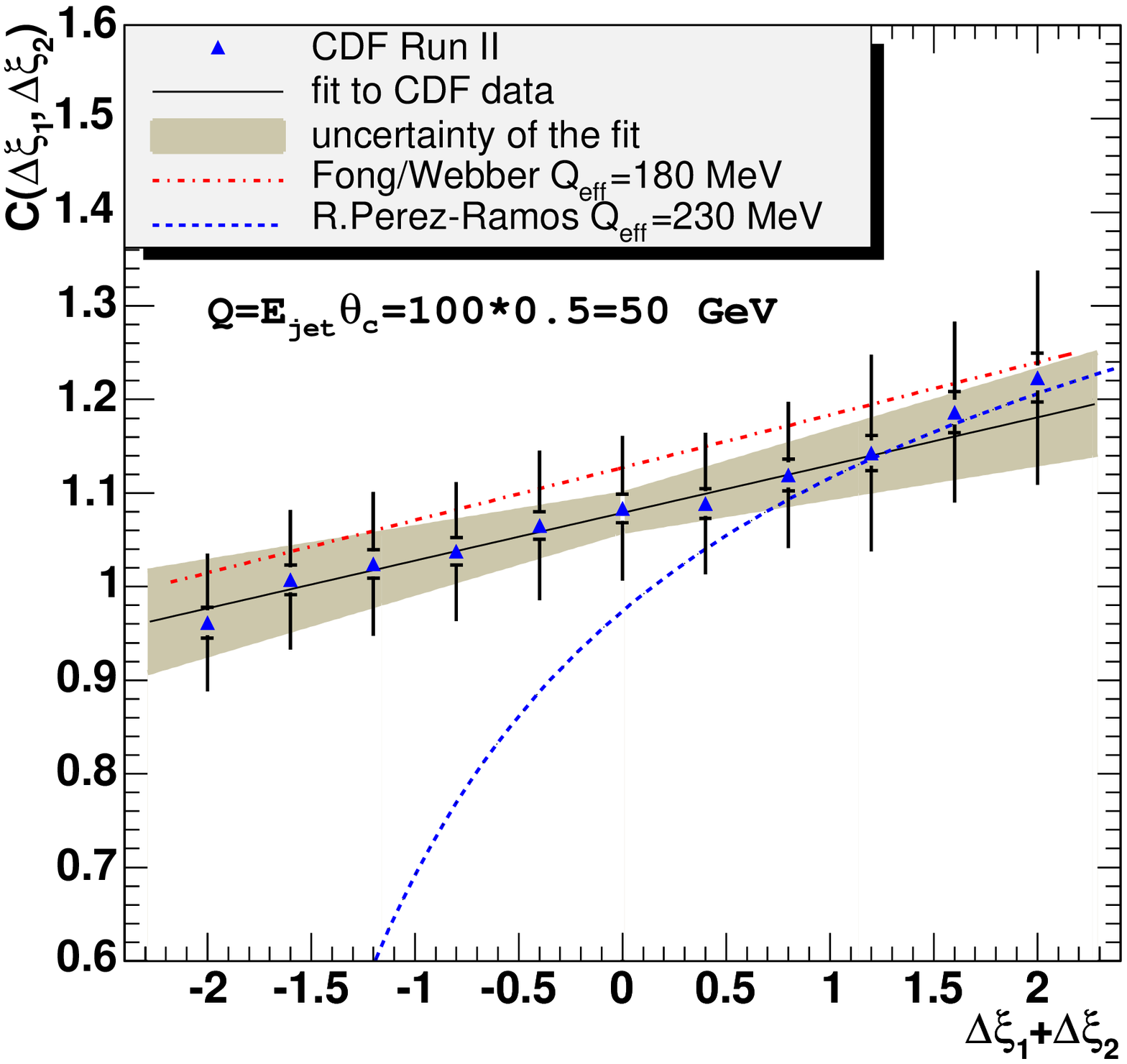}
	\caption{Same as in Fig.~\ref{CorrDataNLLA1} but for $Q=50$ GeV.}
	\label{CorrDataNLLA4}
	\end{figure}

	\begin{figure}
	\includegraphics[width=2.8in]
	{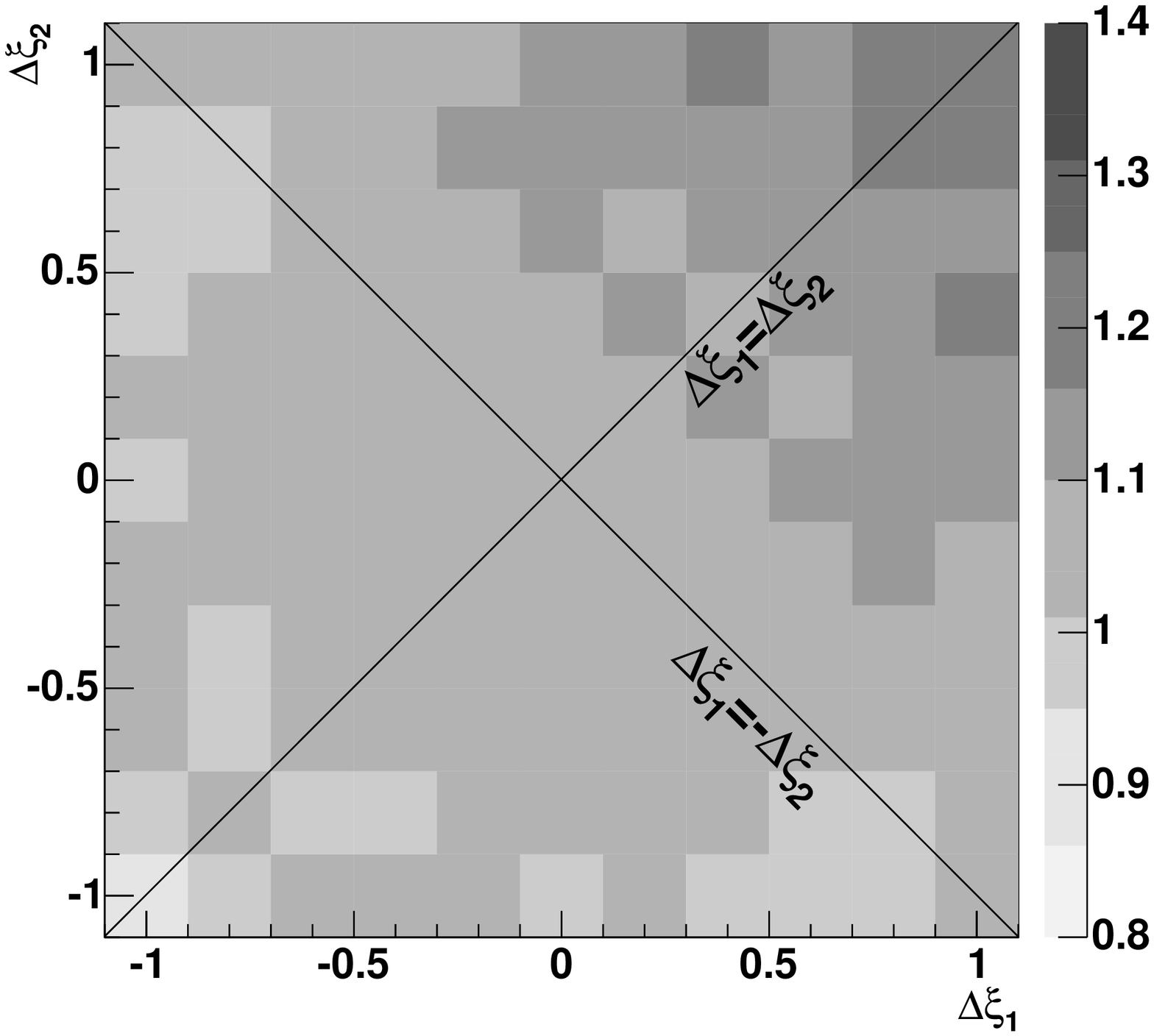}
	\\
	\includegraphics[width=2.8in]
	{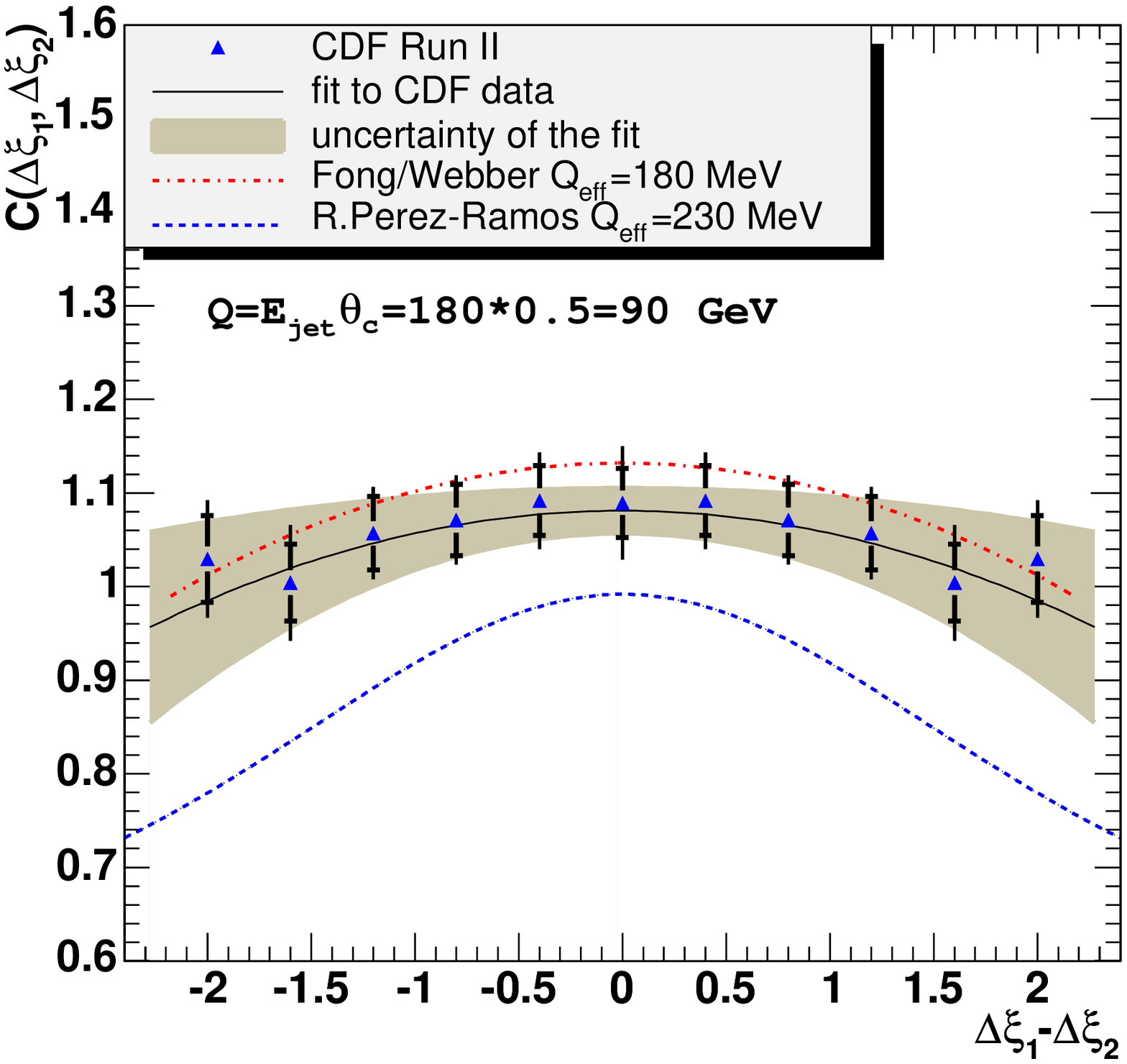}
	\\
	\includegraphics[width=2.8in]
	{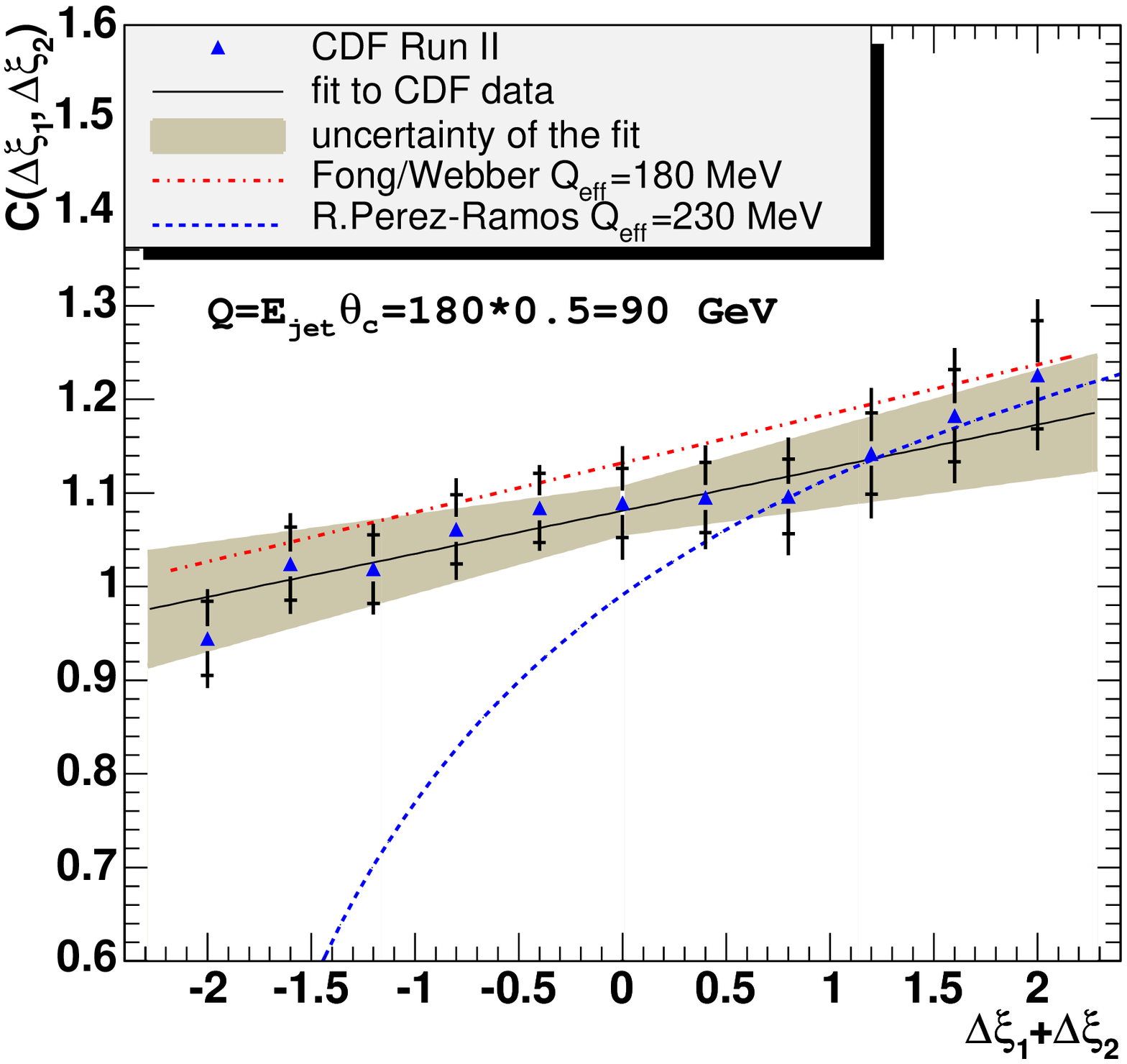}
	\caption{Same as in Fig.~\ref{CorrDataNLLA1} but for $Q=90$ GeV.}
	\label{CorrDataNLLA7}
	\end{figure}

	\begin{figure}
	\includegraphics[width=4.1in]
	{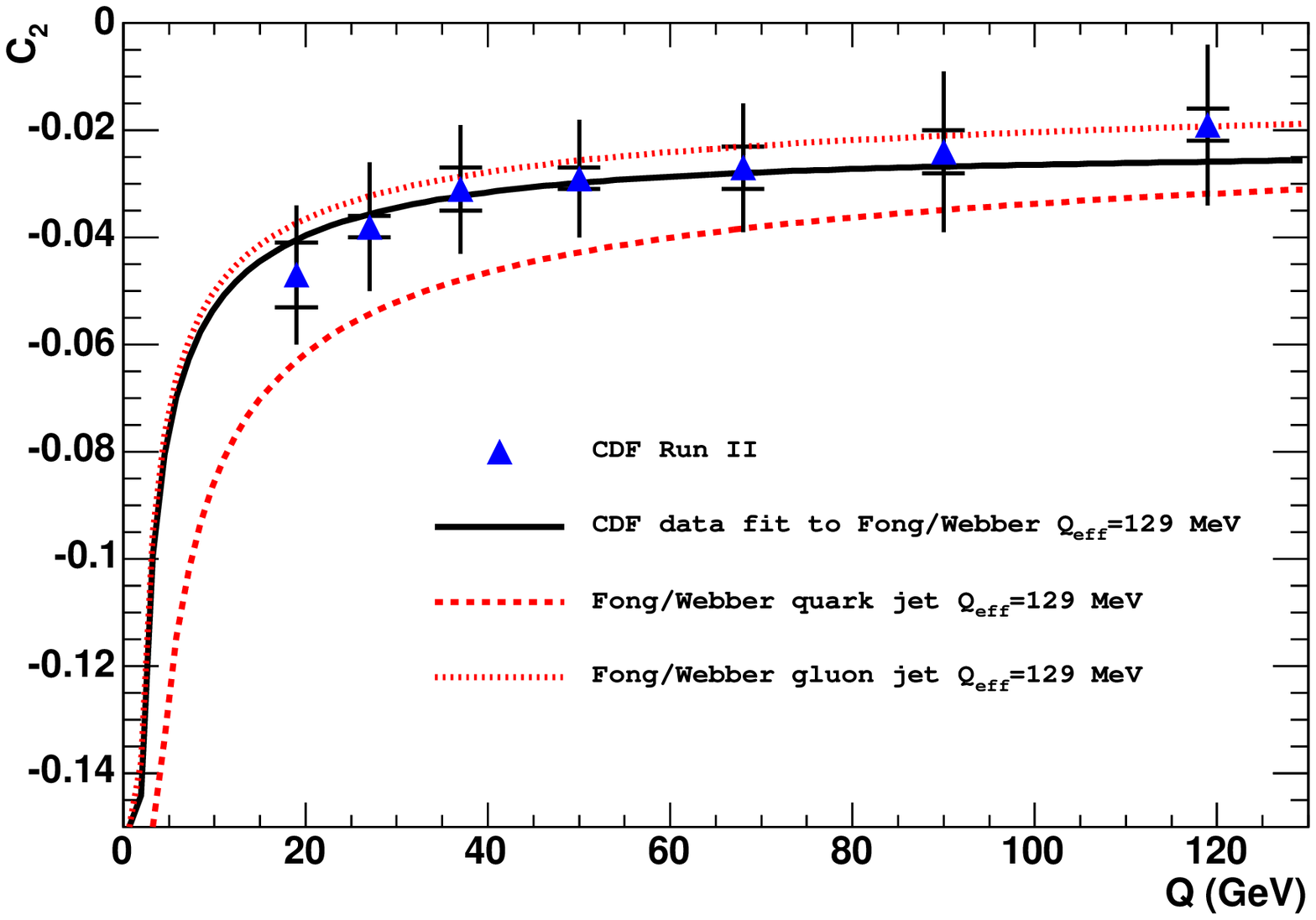}
	\includegraphics[width=4.1in]
	{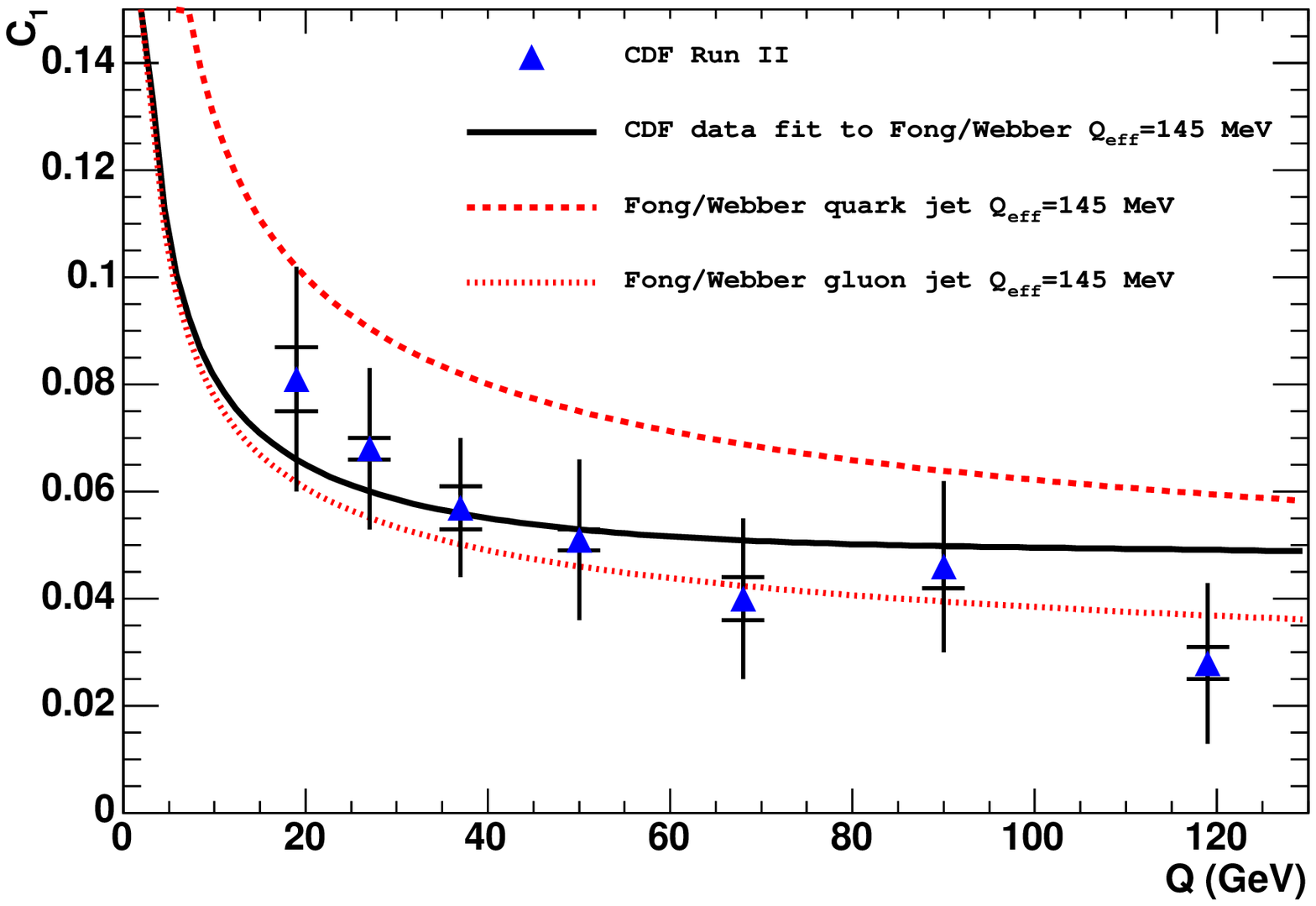}
	\includegraphics[width=4.1in]
	{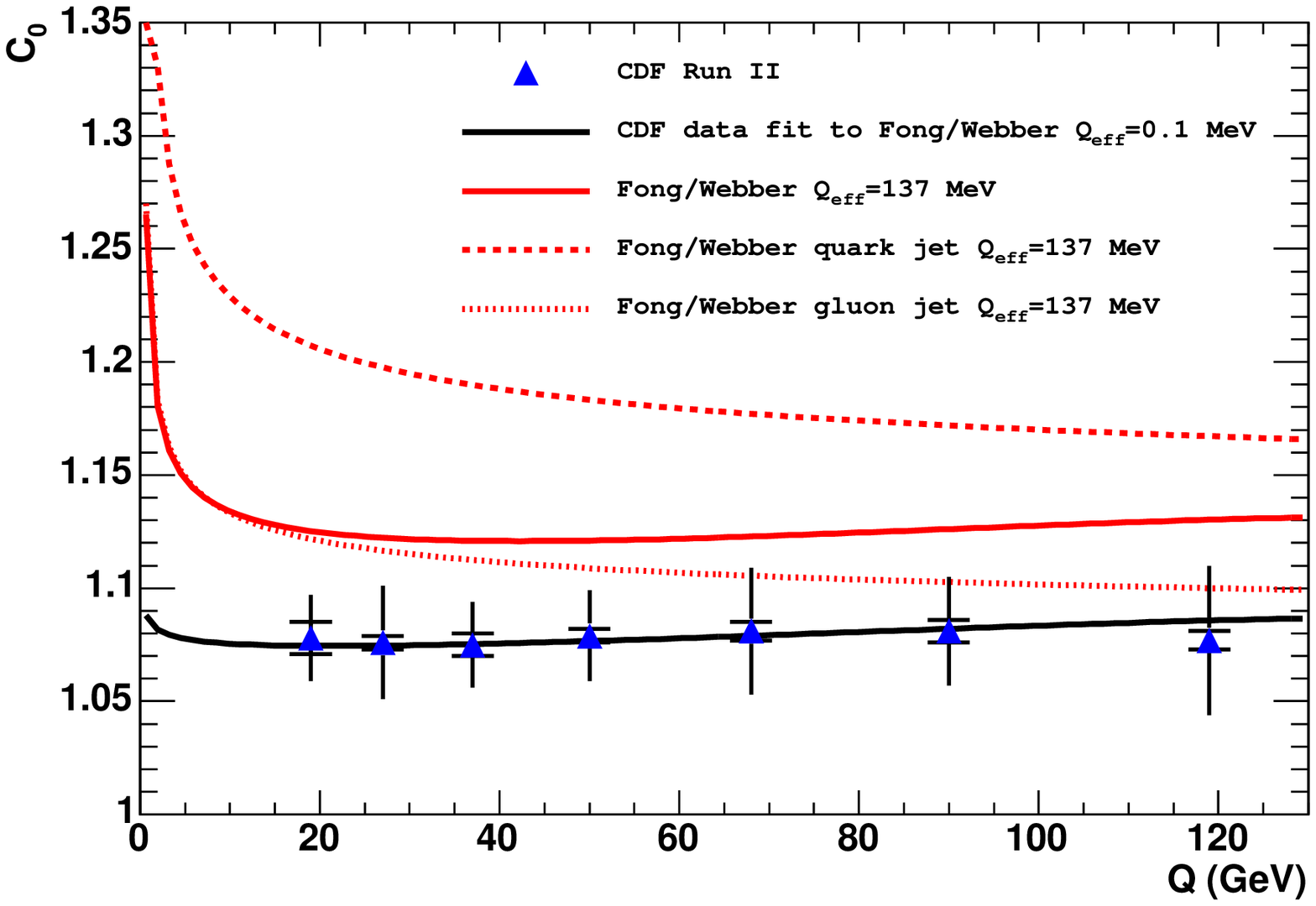}
	\caption{The dependence of correlation parameters $c_{2}$ (top), $c_{1}$ (middle), and $c_{0}$ (bottom) on jet hardness. The Fong-Webber function \cite{Webber} is fit to the CDF data points. The Fong-Webber predictions for pure quark and pure gluon jet samples are also shown.} 
	\label{CorrEvol}
	\end{figure}

	Figure~{\ref{CorrEvol}} shows the dependence of parameters $c_{0}$, $c_{1}$, and $c_{2}$ on jet hardness $Q$. Each data point corresponds to the value of one parameter measured in a particular dijet mass bin. The $c_{0}$ parameter shows almost no dependence on $Q$, while $\left| c_{1} \right|$ and $\left| c_{2} \right|$ decrease with increasing $Q$. This indicates that the correlations are stronger in low energy jets. The distributions are fit to the Fong-Webber function with $Q_\mathit{eff}$ treated as the only free parameter. The fits are represented by solid lines. Theoretical curves for pure quark and gluon jets in the final state are also shown. We use the results of the Fong-Webber calculation {\cite{Webber}} to fit the dependence of these parameters on jet hardness and to extract the parameter $Q_\mathit{eff}$. Results of the Perez-Ramos calculation are not used for the measurement of $Q_\mathit{eff}$ due to the lack of the corresponding analytical expressions.
	The value of $Q_\mathit{eff}$ obtained from the fit of $c_{1}$ is $145 \pm 10\mathrm{(stat)} ^{+79}_{-65} \mathrm{(syst)}$ MeV. The value of $Q_\mathit{eff}$ obtained from the fit of $c_{2}$ is $129\pm12 \mathrm{(stat)}^{+86}_{-71} \mathrm{(syst)}$ MeV. The average value of $Q_\mathit{eff}$ extracted from the combined fit of $c_{1}$ and $c_{2}$ is $137^{+85}_{-69}$ MeV and is consistent with $Q_\mathit{eff}$ extracted from the fits of inclusive particle momentum distributions. The dependence of $c_{0}$ on $Q$ has an offset of $\sim$0.06. This parameter, as opposed to $c_{1}$ and $c_{2}$, is very sensitive to the peak position $\xi_{0}$ of the inclusive momentum distribution. In the data the correlation distributions are measured around the true peak position while in the theoretical calculation of $\xi_{0}$ the unknown constant term $O(1)$ as well as all terms beyond the leading order are neglected. Therefore, theory can control only the dependence of this parameter on energy and not its absolute value. For this reason we exclude $c_{0}$ from the measurement of $Q_\mathit{eff}$. A formal fit of the dependence of $c_{0}$ on $Q$ to the theoretical function gives the value $Q_\mathit{eff}=0.10 \pm 0.08$ MeV. This value, however, does not have physical meaning for the above mentioned reasons. Other than the offset, $c_{0}$ shows very weak, if any, $Q$ dependence, which is consistent with the theory.
	As a cross-check we have measured correlation distributions for pairs of tracks from opposite jets. For our value of the opening angle $\theta_{c}=0.5$ radians, no correlations are observed.

\section{Comparison to Monte Carlo}

	We compare the momentum correlation distributions of charged particles in data to {\sc pythia} tune A and {\sc herwig}~6.5 predictions. Predictions of the two Monte Carlo generators are in good agreement with each other and with results obtained from data. Figures~{\ref{CorrDataMC1}}, {\ref{CorrDataMC4}}, and {\ref{CorrDataMC7}} show the correlation distributions in data compared to {\sc pythia} tune A and {\sc herwig}~6.5 predictions at the level of stable charged hadrons.

	\begin{figure}
	\includegraphics[width=2.8in]
	{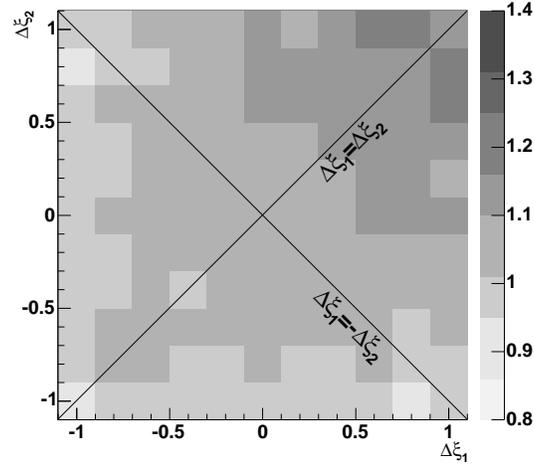}
	\\
	\includegraphics[width=2.8in]
	{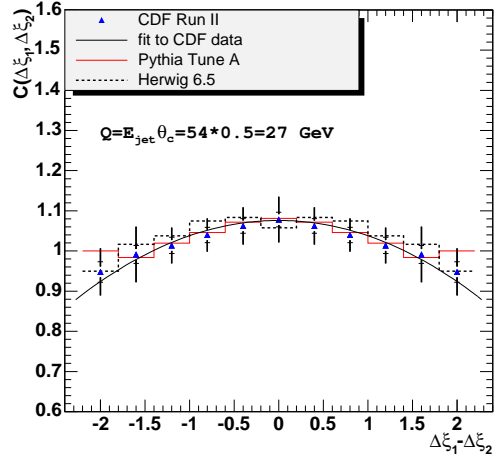}
	\\
	\includegraphics[width=2.8in]
	{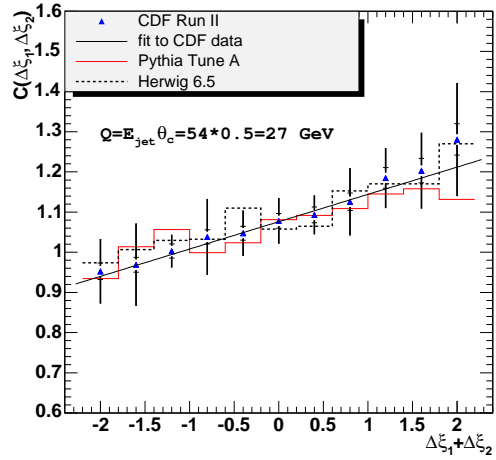}
	\caption{Hadron-level two-particle momentum correlations in jets in the restricted cone of size $\theta_{c}=0.5$ radians for the dijet mass bin with $Q=27$ GeV using {\sc pythia} tune A (top). Data correlations are compared to the hadron momentum correlations using the  {\sc pythia} tune A and {\sc herwig}~6.5 event generators. Central diagonal profiles $\Delta\xi_{1}=-\Delta\xi_{2}$ (middle) and $\Delta\xi_{1}=\Delta\xi_{2}$ (bottom) of the distributions are shown.}
	\label{CorrDataMC1}
	\end{figure}

	\begin{figure}
	\includegraphics[width=2.8in]
	{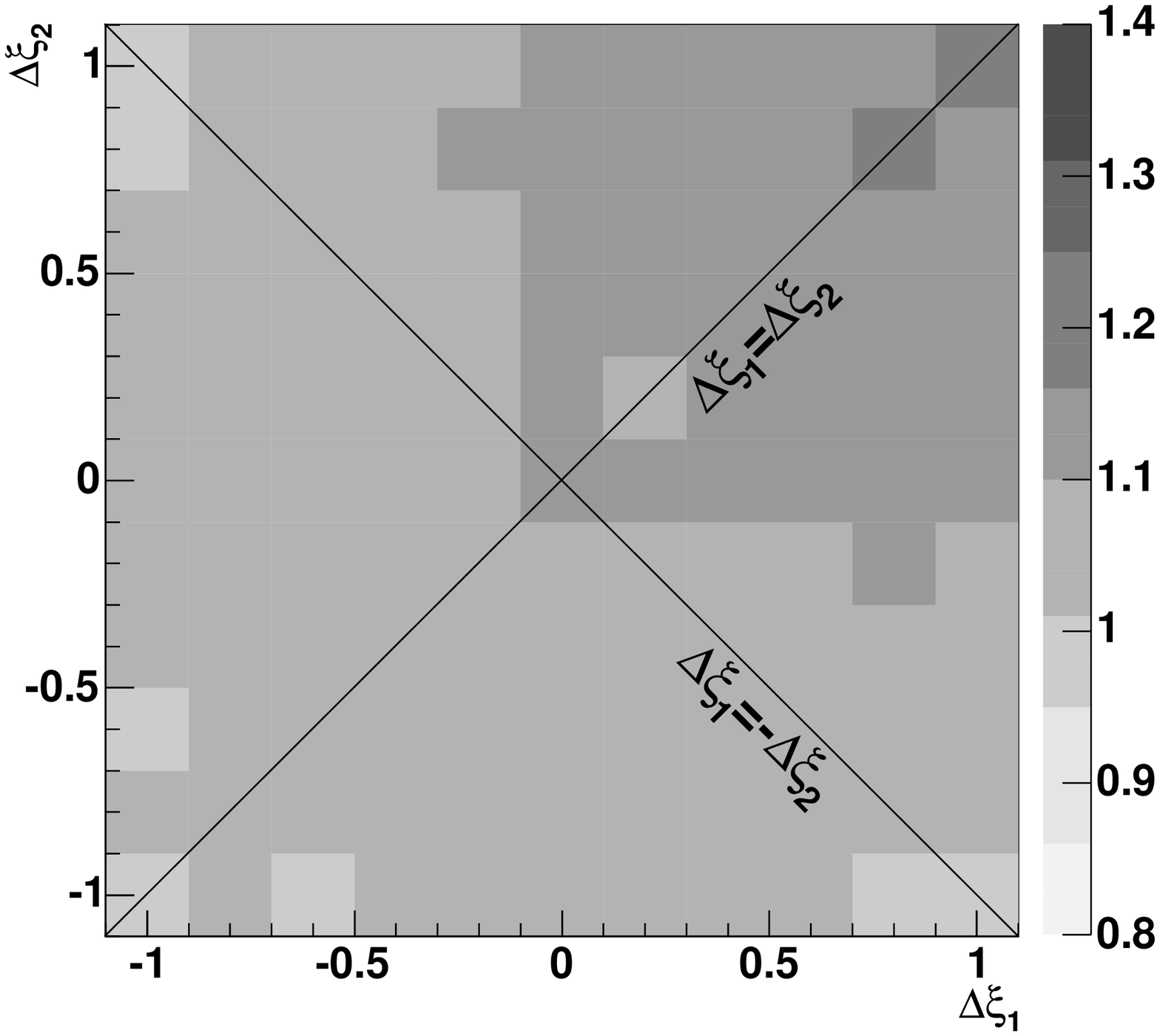}
	\\
	\includegraphics[width=2.8in]
	{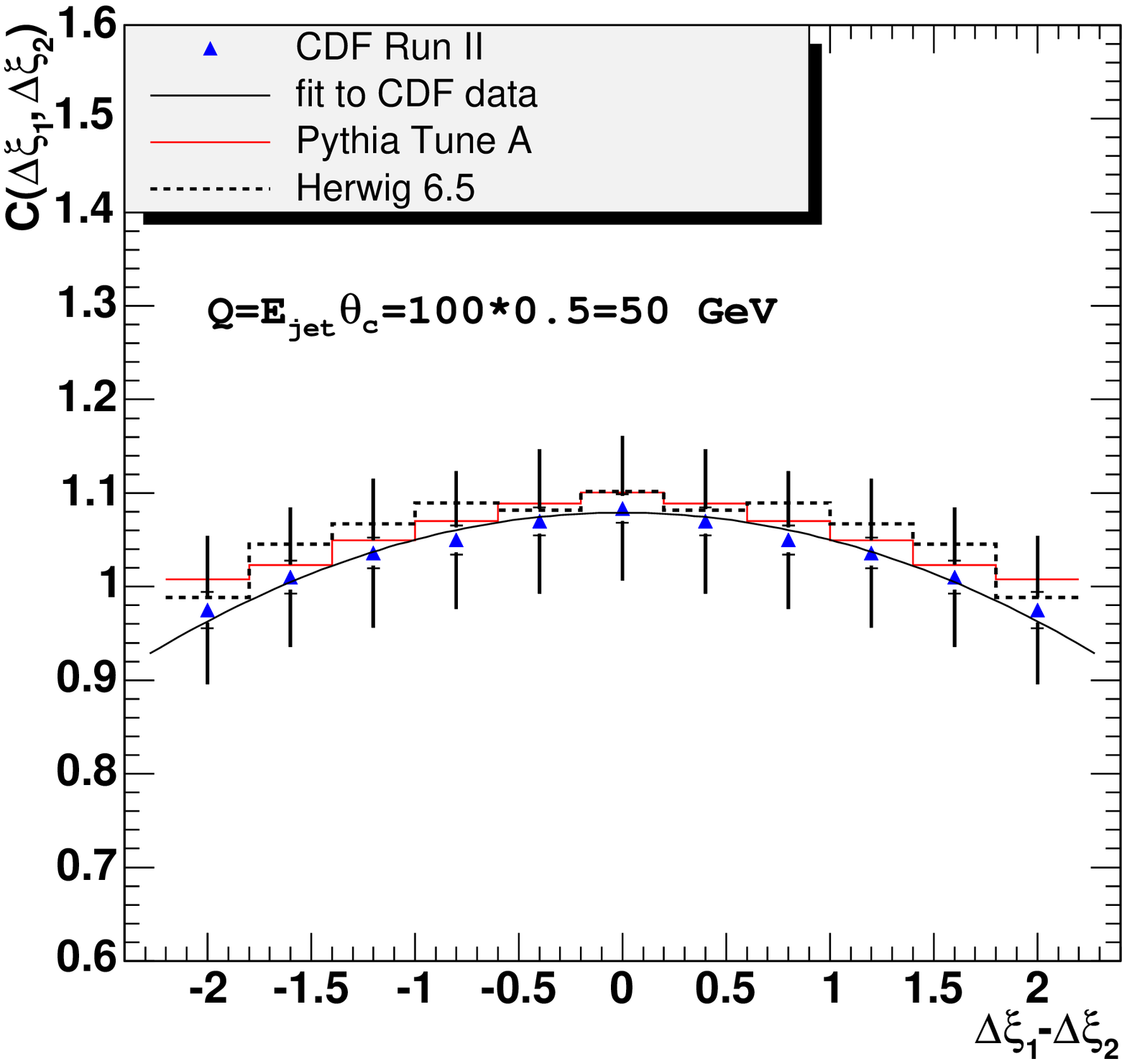}
	\\
	\includegraphics[width=2.8in]
	{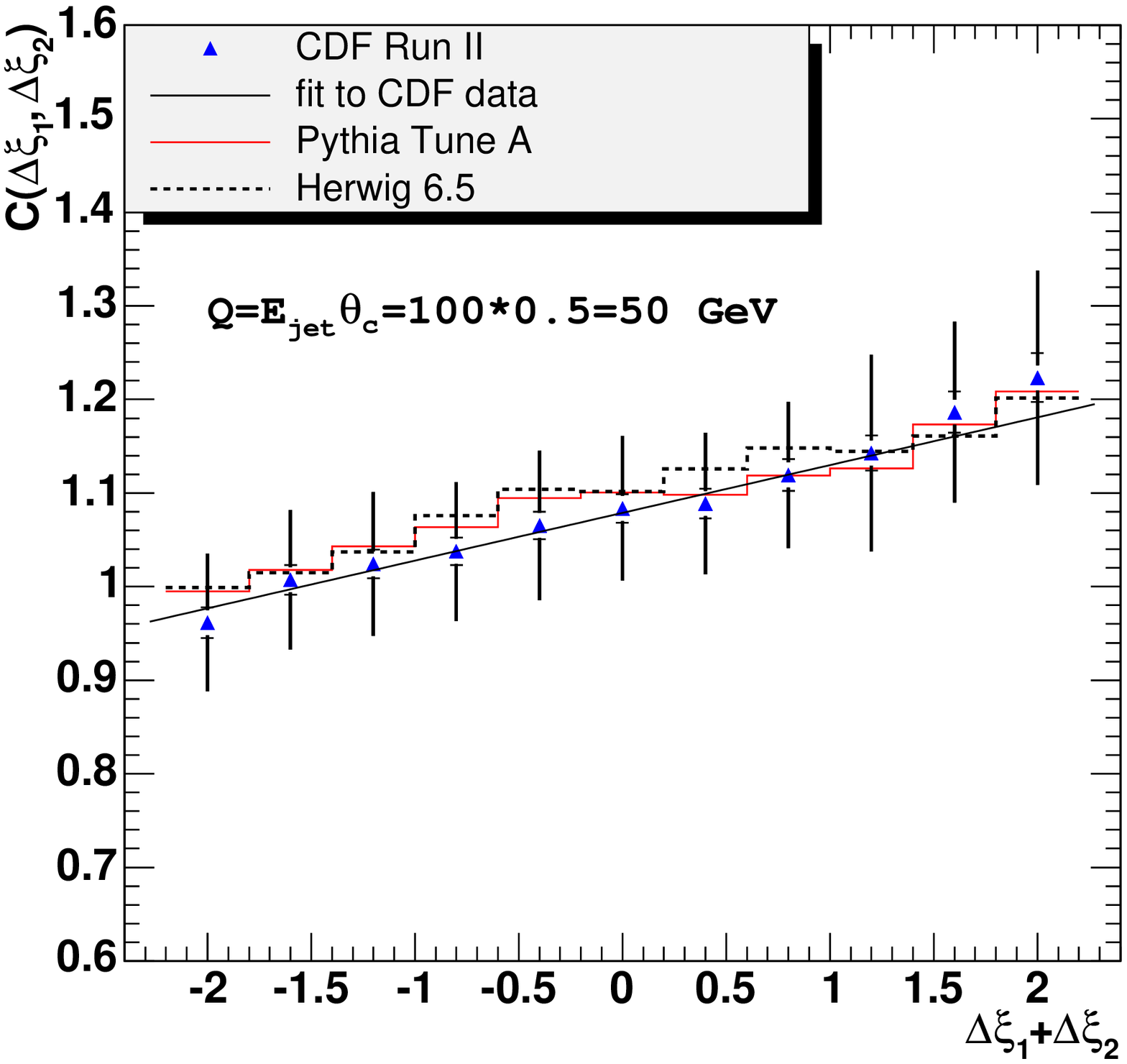}
	\caption{Same as in Fig.~\ref{CorrDataMC1} but for $Q=50$ GeV.}
	\label{CorrDataMC4}
	\end{figure}

	\begin{figure}
	\includegraphics[width=2.8in]
	{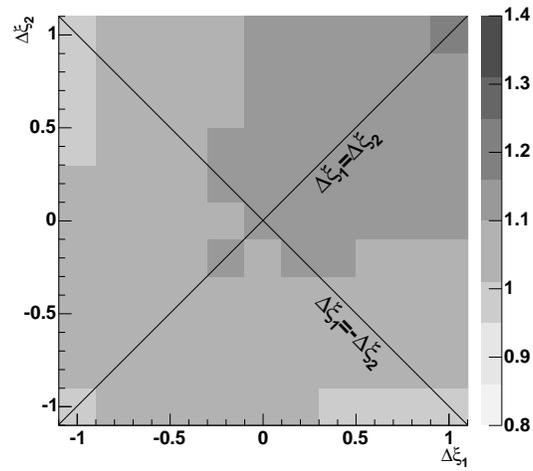}
	\\
	\includegraphics[width=2.8in]
	{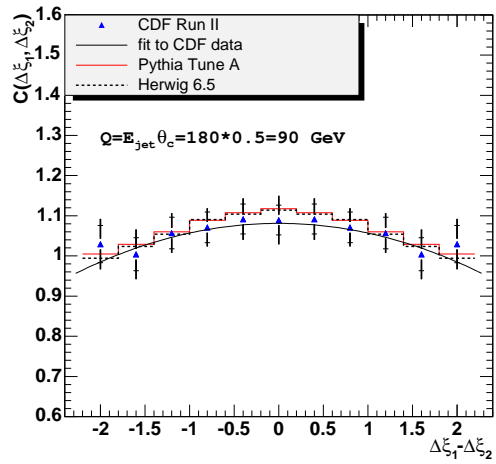}
	\\
	\includegraphics[width=2.8in]
	{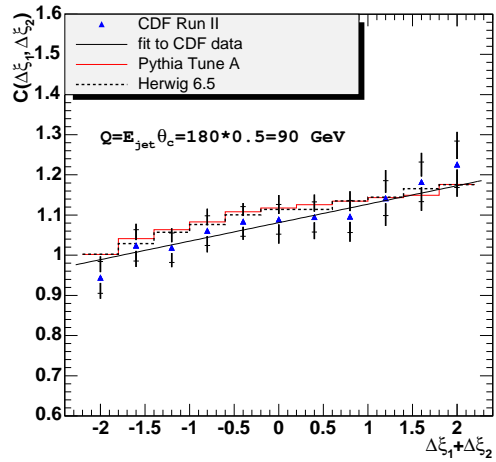}
	\caption{Same as in Fig.~\ref{CorrDataMC1} but for $Q=90$ GeV.}
	\label{CorrDataMC7}
	\end{figure}

	\section{Summary}
	The two-particle momentum correlation distributions of charged particles in jets from dijet events have been measured over a wide range of dijet masses from 66 to 563 GeV/c$^{2}$. The jets were produced in $p\bar p$ collisions at a center of mass energy of 1.96 TeV. The measurements have been performed for particles in a restricted cone around the jet direction with opening angle $\theta_{c}=0.5$ radians.

	The data are compared to the next-to-leading log approximation calculations combined with the hypothesis of local parton-hadron duality (LPHD). Overall, the data and the theory show the same trends over the entire range of dijet energies. The parton shower cutoff scale $Q_\mathit{eff}$ is set equal to $\Lambda_\mathit{QCD}$ and is extracted from fits of the dependence of the correlation parameters, $c_{1}$ and $c_{2}$, defining the strength of the correlation, on jet hardness $Q$. The average value of $Q_\mathit{eff}$ extracted from the combined fit of $c_{1}$ and $c_{2}$ is $137^{+85}_{-69}$ MeV and is consistent with $Q_\mathit{eff}$ extracted from the fits of inclusive particle momentum distributions and with the results of a previous CDF measurement \cite{SafonovPRD}. As predicted, the parameter $c_{0}$ shows little, if any, dependence on jet energy; however, we observe a substantial systematic offset between the experimental and theoretical values. The parameter $c_{0}$ is excluded from the measurement of $Q_\mathit{eff}$ because of its large theoretical uncertainty. The modified leading log approximation predictions qualitatively show the same trends; however, the quantitative disagreement with the data is obviously larger in this case.

	The {\sc pythia} tune A and {\sc herwig}~6.5 Monte Carlo event generators are found to reproduce the correlations in data fairly well.

	The results of this analysis indicate that the parton momentum correlations do survive the hadronization stage of jet fragmentation, giving further support to the hypothesis of LPHD.

\section{Acknowledgments}

	The authors are very grateful to R.~Perez-Ramos for collaborative work and to F.~Arleo, B.~Machet, and Yu.~Dokshitzer for a number of very fruitful discussions. We thank the Fermilab staff and the technical staffs of the participating institutions for their vital contributions. This work was supported by the U.S. Department of Energy and National Science Foundation; the Italian Istituto Nazionale di Fisica Nucleare; the Ministry of Education, Culture, Sports, Science and Technology of Japan; the Natural Sciences and Engineering Research Council of Canada; the National Science Council of the Republic of China; the Swiss National Science Foundation; the A.P. Sloan Foundation; the Bundesministerium f\"ur Bildung und Forschung, Germany; the Korean Science and Engineering Foundation and the Korean Research Foundation; the Science and Technology Facilities Council and the Royal Society, UK; the Institut National de Physique Nucleaire et Physique des Particules/CNRS; the Russian Foundation for Basic Research; the Comisi\'on Interministerial de Ciencia y Tecnolog\'{\i}a, Spain; the European Community's Human Potential Programme; the Slovak R\&D Agency; and the Academy of Finland.

\newpage


\clearpage

\bibliography{basename of .bib file}

\begin{references}
\bibitem{TrigTower} A trigger tower is defined by adding the information of the two adjacent calorimeter towers.
\bibitem{missingEt} \met ~is defined as the norm of $-\Sigma_{i} E_{T}^{i} \cdot \vec{n_{i}}$, where $\vec{n_{i}}$ is the unit vector in the azimuthal plane that points from the beamline to the i-th calorimeter tower.
\end{references}

\end{document}